\documentclass[12pt,letterpaper]{article}
\usepackage{epsfig,rotating,setspace,latexsym,amsmath,epsf,amssymb,amsfonts,bm,theorem,cite,caption,subcaption,enumerate,longtable,accents}
\usepackage{algorithm,algorithmic,graphicx,epsf,authblk,epstopdf,url,color,multirow}
\usepackage{mathtools,comment}

\newtheorem{theorem}{Theorem}

\newtheorem{remark}{Remark}

\newtheorem{example}{Example}
\newenvironment{Proof}[1]{\medskip\par\noindent{\bf Proof:\,}\,#1}{{\mbox{\,$\blacksquare$}\par}}

\allowdisplaybreaks

\title{Private Federated Submodel Learning via \\ Private Set Union\thanks{This work was supported by ARO Grant W911NF2010142.}}

\author{Zhusheng Wang \qquad Sennur Ulukus\\
	\normalsize Department of Electrical and Computer Engineering\\
	\normalsize University of Maryland, College Park, MD 20742\\
	\normalsize  \emph{zhusheng@umd.edu} \qquad \emph{ulukus@umd.edu}}

\begin{document}
\date{}
\maketitle

\vspace*{-1.0cm}

\begin{abstract}
We consider the federated submodel learning (FSL) problem and propose an approach where clients are able to update the central model information theoretically privately. Our approach is based on private set union (PSU), which is further based on multi-message symmetric private information retrieval (MM-SPIR). The server has two non-colluding databases which keep the model in a replicated manner. With our scheme, the server does not get to learn anything further than the subset of submodels updated by the clients: the server does not get to know which client updated which submodel(s), or anything about the local client data. In comparison to the state-of-the-art private FSL schemes of Jia-Jafar and Vithana-Ulukus, our scheme does not require noisy storage of the model at the databases; and in comparison to the secure aggregation scheme of Zhao-Sun, our scheme does not require pre-distribution of client-side common randomness, instead, our scheme creates the required client-side common randomness via random symmetric private information retrieval (RSPIR) and one-time pads. Our system is initialized with a replicated storage of submodels and a  sufficient amount of common randomness at the two databases on the server-side. The protocol starts with a common randomness generation (CRG) phase where the two databases establish common randomness at the client-side using RSPIR and one-time pads (this phase is called FSL-CRG). Next, the clients utilize the established client-side common randomness to have the server determine privately the union of indices of submodels to be updated collectively by the clients (this phase is called FSL-PSU). Then, the two databases broadcast the current versions of the submodels in the set union to clients. The clients update the submodels based on their local training data. Finally, the clients use a variation of FSL-PSU to write the updates back to the databases privately (this phase is called FSL-write). As the databases at the server do not communicate, as a novel approach, we utilize randomly chosen alive clients to route the required information between the two databases. Our proposed private FSL scheme is robust against client drop-outs, client late-arrivals, and database drop-outs.
\end{abstract}

\section{Introduction}
A standard machine learning approach requires the whole training data to be centralized on a machine or a server to learn the target model. A practical challenge in today's machine learning applications is that the training data is distributed over multiple isolated clients, e.g., training a recommendation system in e-commerce requires interaction with a large number of mobile devices held by different customers in order to employ their data. Due to the distributed nature of training data, a new approach called federated learning (FL) has been proposed \cite{FL,FL_Yangconcept}. In FL, each client performs a round of local training by using its own data, after being selected and having obtained the current model from the global server. In an FL setting, collaboratively performing a learning task while protecting the privacy of the local data stored in each client against the global server is a significant issue. An intuitive way for FL to achieve privacy is to use a secure aggregation protocol such that no individual client's update can be inspected by the global server \cite{SecAgg}. As a stand-alone topic, secure aggregation has been a continuously active topic in the computer science literature, see \cite{SecAgg,SecAgg+,TurboAgg,FastSecAgg}. Recently, information theoretically secure aggregation schemes towards achieving optimal communication cost have been proposed for various common randomness distribution settings among the clients, see \cite{IT_SecAgg,IT_SecAgg_UGKey,IT_SecAgg_Region}. However, in these papers, the communication costs of the input and the common randomness are considered separately, without explicitly stating the common randomness generation and allocation process in a concrete realization of a scheme. Another information theoretic secure aggregation scheme is investigated in \cite{LightSecAgg} where the common randomness sharing relies on the private communication links between each two clients.

As an extension of the now well-established FL framework, recently, a new framework called federated submodel learning (FSL) has been put forward to further reduce the communication and computation overhead at both server and client sides \cite{FSL}. In the submodel framework, the full learning model stored in the server is divided into multiple submodels based on their data characteristics. Each selected client downloads only the needed submodels from the server and then uploads the corresponding submodel updates according to the type of their local data. As pointed out by \cite{FSL}, there are two fundamental problems that can be abstracted out of the FSL framework: One is how can each client download its desired submodels from the curious server without revealing the indices of these submodels to the server. The other is how can each client update these desired submodels still without revealing the indices or the content of the updated submodels to the curious server. The first one is a \emph{private read} problem, and the second one is a \emph{private write} problem.

Reference \cite{FSL} proposes a weak-privacy approach as follows: First, the server attains the union of desired submodel indices from all the selected clients in a secure manner. As introduced in \cite{KS05,PSU}, this is basically a multi-party private set union (MP-PSU) problem and it is resolved through a Bloom filter in \cite{FSL}. Then, the server delivers this inaccurate union result which may have false positives to each selected client. Once each client receives this information, it randomizes its real desired submodel index set within the scope of this union through a randomized response. After receiving the submodels corresponding to this randomized desired submodel index set from the server, each client trains only the submodels whose indices are within the intersection of its real and randomized desired submodel index sets, and then uploads the updates through secure aggregation. Thus, \cite{FSL} partially solves the above two fundamental privacy problems, while sacrificing the update efficiency of clients. In addition, the secure aggregation problem in FSL is related to the private multi-group aggregation problem in \cite{PMG_Agg}, with different submodels viewed as distinct groups.

A strong-privacy FSL approach is introduced in \cite{XSTFSL} based on cross subspace alignment \cite{XSTPIR}. In this approach, only one client who is interested in a specific submodel participates in one round of FSL process. If each database stores the plain full model for learning, then databases can learn the updates made by this client by comparing the stored information before and after training. Thus, databases store the full model across multiple distributed databases in the server in a noisy manner based on a threshold secret sharing scheme. This storage redundancy helps hide the index and the content of the written (updated) submodel from databases; that is, databases cannot tell which submodel has been updated and what the updated value is. This storage redundancy also helps secure the submodels against the databases; that is, even though the databases hold the submodels, they cannot know the actual values of the submodels. Concurrently, an improved scheme in terms of communication cost efficiency is given in \cite{Sajani_FSL1}, and extended to the case of sparsified updates which further reduces the communication cost \cite{Sajani_FSL_Trans}.

In this paper, we propose a new FSL scheme that retains the main advantages of the above-mentioned two approaches with a privacy protection level that is in between. In particular, we propose a two-phase scheme. In the first phase, the server securely calculates the union of the clients' desired submodel indices, and in the second phase, the server securely aggregates clients' generated updates in the calculated set union. The first phase is well-known as the private set union (PSU) problem and referred to as FSL-PSU phase in this paper. In this phase, the server learns only the union of the submodel indices to be updated by the clients, but does not learn the submodel indices to be updated by individual clients. The second phase is well-known as the secure aggregation problem and referred to as FSL-write phase in this paper. In this phase, the server securely aggregates the submodel updates without being able to inspect individual updates. That is, in both phases, the server can only learn the ultimate result, without being able to know which client has made which contribution to the ultimate result. We solve these two problems together using two different forms of a common core idea. To that end, we first propose a novel PSU method that is information theoretically secure for the FSL-PSU phase, and modify it to obtain an information theoretically secure aggregation method for the FSL-write phase.

In the field of cryptography, private information retrieval (PIR) refers to a fundamental problem where a user wishes to retrieve a specific message out of a set of messages that is stored across multiple non-colluding and replicated databases while completely concealing the index of the desired message from each individual database \cite{PIR_ORI}. Symmetric PIR (SPIR) additionally requires that the user is not able to obtain any knowledge about the remaining messages in the databases after downloading its desired message \cite{SPIR_ORI}. Following the seminal paper that focuses on the information theoretic capacity of multi-database PIR \cite{PIR}, PIR and SPIR have attracted a tremendous amount of attention in the field of information theory recently, e.g., \cite{PIR_coded, ColludingPIR, BPIRjournal, PIR_Eavesdropper, MM-PIR, tandon_cache_2017, PrefetchingPIR, PIR_PSI, StorageConstrainedPIR, StorageCost, XSTPIR, ProbPIR, AleakyPIR, SemanticPIR, SPIR, SPIR_atPIR, CommCost_ISIT2022, SPIR_Eavesdropper, SPIR_coded, SPIR_Collusion, PSI_journal}. As a non-trivial variation of SPIR, in multi-message SPIR (MM-SPIR), the user wishes to retrieve multiple messages at a time \cite{PSI_journal}. The paper \cite{PSI_journal} also establishes the equivalence between the MM-SPIR problem and the private set intersection (PSI) problem. In the PSI problem, two parties wish to determine the common elements in their possessed data sets without leaking any further information about the remaining elements. Note that the constraints in the PSI and PSU problems are analogous. In this paper, we establish the equivalence between the PSU and MM-SPIR problems. 

\begin{figure}[t]
\centering
\includegraphics[width=0.8\linewidth]{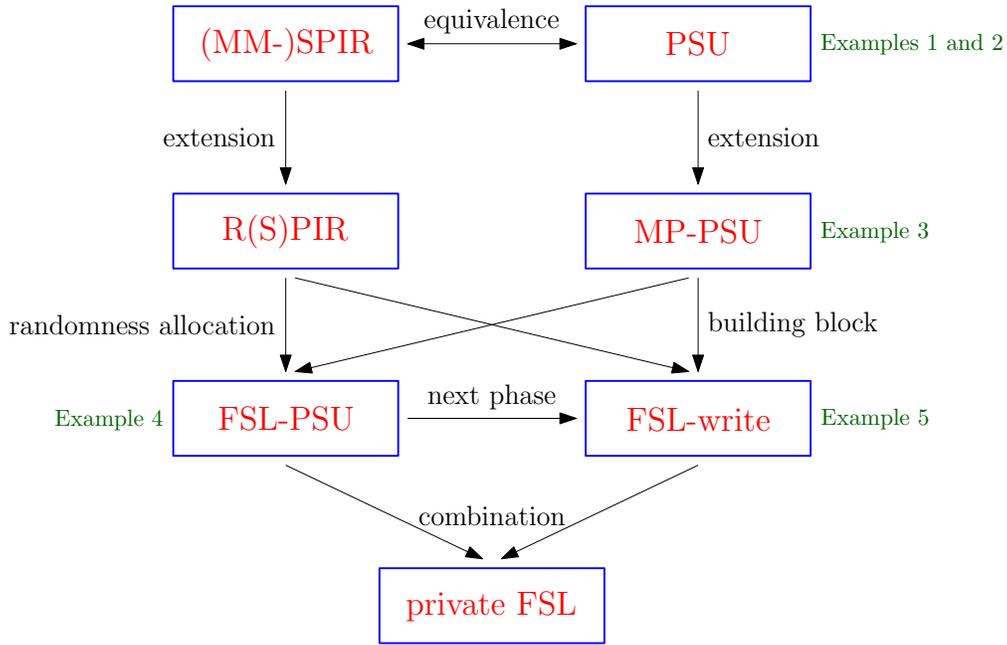}
\caption{Techniques used, their relationships, and the roadmap of the development of the private FSL in this paper. Example references next to boxes show examples presented later in the paper to describe the functionality of each box.}
\label{Roadmap}
\end{figure}

In this work, similar to typical PIR/SPIR formulations, we consider the setting where the FSL server has multiple databases, and focus on the simplest such setting where the server has two databases. In Fig.~\ref{Roadmap}, we show the techniques used, their relationships, and the roadmap of the development in this paper. The references next to boxes in Fig.~\ref{Roadmap} show examples presented later in the paper to describe the functionality of each box. The classical information theoretic SPIR serves as a starting point to formulate our new FSL achievable scheme. We first show the equivalence between the MM-SPIR and PSU problems, and then extend it to the case of muti-party PSU (MP-PSU) as in \cite{PSI_journal,MP-PSI_journal}; thus, we use an MM-SPIR scheme to construct an information theoretic achievable scheme for MP-PSU. We use this MP-PSU framework as a building block to develop FSL-PSU and FSL-write phases jointly noting that FSL-PSU and FSL-write are fundamentally similar security problems. For the FSL-PSU and FSL-write phases, an amount of necessary common randomness unknown to each individual database needs to be shared among the clients. Still starting from the SPIR problem, we take advantage of random SPIR (RSPIR) that is newly formulated in \cite{RSPIR} to accomplish the common randomness allocation across the clients.

A classical multiple-database secure aggregation scheme is proposed in \cite{Prio} whose core idea is outlined next: In a setting with a server consisting of $N$ independent databases, and $C$ individual clients, each client $i$ splits its input $z^{\langle i \rangle}$ into $N$ shares\footnote{For a positive integer $Z$, we adopt the notation $[Z] = \{1, 2, \dots , Z\}$ in this work for simplicity. Therefore, the $[N]$ here and $[C], [K], [N_1], [L]$ later are all used to denote a set of integers.} $z^{\langle i \rangle}_{[N]} = \{z^{\langle i \rangle}_{1},\dots,z^{\langle i \rangle}_{N}\}$ such that the sum of these shares exactly equals the input, i.e., $\sum_{j \in [N]} z^{\langle i \rangle}_{j} = z^{\langle i \rangle}$. Subsequently, each client $i$ sends the share $z^{\langle i \rangle}_{j}$ to database $j$ for all $j \in [N]$. After collecting shares from all the clients, each database publishes the sum of its received shares $\sum_{i \in [C]} z^{\langle i \rangle}_{j}$. Then, the server computes the sum of all the inputs possessed by the clients $\sum_{j \in [N]} \sum_{i \in [C]} z^{\langle i \rangle}_{j} = \sum_{i \in [C]} \sum_{j \in [N]} z^{\langle i \rangle}_{j} = \sum_{i \in [C]} z^{\langle i \rangle}$. Although this simple scheme achieves the required privacy in that the databases can only obtain a sum and nothing beyond that in terms of each client's private input, it has some unavoidable disadvantages. First, the upload cost is high as each client needs to upload an answer to each database. Second, once one of these databases cannot function normally, the server can only receive some random values that is far from the ultimate sum. Third, the communication among the databases in the server is required.  

In this paper, we propose a new achievable scheme for private distributed FSL primarily through unifying the FSL-PSU and FSL-write problems in the same framework. In a practical FL implementation, communication among the clients is generally unstable (if any) as the index set of the selected clients keeps on changing in each round, and it is extremely expensive to establish direct communication channels among the clients. Since the inter-database communication is not permitted either, our scheme is designed to rely only on truthful communication between the clients and the databases through a number of direct secure and authenticated client-database communication channels. Further, due to the long duration of the FSL-write phase, it is possible for some clients to drop out as this phase moves on. Thus, we design our scheme in such a way that even if some clients lose their connection to the server, our scheme continues to work normally. Our scheme is also robust against client drop-outs during the FSL-PSU phase. In addition, it is possible that some clients' generated answers arrive at their associated databases late and the corresponding databases make the wrong judgement that the clients have dropped out. Our scheme is designed such that these late answers do not leak any additional information about these late clients to the databases; this is referred to as robustness against client late-arrivals. Moreover, our scheme continues to work normally even when some of the databases become inactive, especially when the total number of databases is large enough. Another critical aspect of our scheme is that it accounts for the total communication cost incurred, including the cost of common randomness generation at the clients. Finally, our FSL scheme can be run in an iterative fashion in multiple rounds of FSL process, until a predefined termination criterion is met.

\section{Problem Formulation} \label{Problem Formulation}
\subsection{MM-SPIR} \label{MM-SPIR Formulation}
As in \cite{PSI_journal}, we consider $N \geq 1$ non-colluding databases with each individual database storing the replicated set of $K \geq 2$ i.i.d.~messages $W_{[K]}=\{W_1,\dots,W_K\}$. The $L$ i.i.d.~symbols within each message are uniformly selected from a sufficiently large finite field $\mathbb{F}_q$, hence,
\begin{align}
    H(W_k) &= L, \quad \forall k \label{Message Length} \\
    H(W_{[K]}) &= H(W_1) + \dots + H(W_K)  = KL \label{Message IID}
\end{align}

The goal of the MM-SPIR problem is to retrieve a set of messages $W_{\Omega}$ out of the message set $W_{[K]}$ without leaking any information regarding the retrieved index set $\Omega =\{i_1,i_2,\cdots,i_P\} \subseteq [K]$ with cardinality $|\Omega|=P$ to any individual database (user privacy constraint), and while not obtaining any further information beyond the desired message set $W_{\Omega}$ (database privacy constraint). The cardinality of the retrieved message set $P$ is public knowledge and known by all the databases. Due to the database privacy constraint, databases need to share some amount of server-side common randomness $\mathcal{R}_S$ that is unknown to the user. The server-side common randomness $\mathcal{R}_S$ is independent of the message set $W_{[K]}$ in the server.

The desired message index set $\Omega$ is a random variable corresponding to a uniform selection of elements without replacement from the set $[K]$ and the sample space of $\Omega$ is the power set of $[K]$. We use $\mathcal{P}$ to denote the realization of the random variable $\Omega$. Based on the desired message set $\Omega$, the user generates a set of queries $Q^{[\Omega]}_{[N]}$ without knowing the message set $W_{[K]}$ stored in the databases, hence,
\begin{align} \label{Message Set Independence}
    I(W_{[K]};Q^{[\Omega]}_{[N]},\Omega) = 0
\end{align} 

For any desired message index set $\mathcal{P}$, after receiving a query from the user, each database responds with a truthful answer based on the stored message set and the server-side common randomness, 
\begin{align} \label{Deterministic Answer} 
   \text{[MM-SPIR deterministic answer]} \quad  H(A_n^{[\mathcal{P}]}|Q_n^{[\mathcal{P}]},W_{[K]},\mathcal{R}_S) = 0, \quad \forall n, ~\forall \mathcal{P} 
\end{align}

Subsequently, the user should be able to decode the desired set of messages reliably after collecting $N$ answers from all the databases, 
\begin{align} \label{Reliability} 
  \text{[MM-SPIR reliability]} \quad &H(W_{\mathcal{P}}|Q_{[N]}^{[\mathcal{P}]},A_{[N]}^{[\mathcal{P}]},\mathcal{P}) = 0, \quad \forall \mathcal{P} 
\end{align}

Due to the user privacy constraint, the query generated to retrieve the desired set of messages should be statistically indistinguishable from other queries. Thus, for all realizations $\mathcal{P}$ and $\mathcal{P}^\prime$, such that $\mathcal{P} \neq \mathcal{P}^\prime$ and $|\mathcal{P}| = |\mathcal{P}^\prime| = P$,
\begin{align} \label{User Privacy}
    \text{[MM-SPIR user privacy]} \quad  (Q_n^{[\mathcal{P}]},A_n^{[\mathcal{P}]},W_{[K]},\mathcal{R}_S) \sim  (Q_n^{[\mathcal{P}^\prime]},A_n^{[\mathcal{P}^\prime]},W_{[K]},\mathcal{R}_S) 
\end{align}
which is equivalent to the following one,
\begin{align}
    \text{[MM-SPIR user privacy]} \quad I(\Omega;Q_n^{[\Omega]},A_n^{[\Omega]},W_{[K]},\mathcal{R}_S) = 0, \quad \forall n
\end{align}

Due to the database privacy constraint, the user should learn nothing about  $W_{\bar{\mathcal{P}}}$ which is the complement of $W_{\mathcal{P}}$, i.e., $W_{\bar{\mathcal{P}}} = W_{[K] \backslash \mathcal{P}}$, 
\begin{align}
    \text{[MM-SPIR database privacy]} \quad I(W_{\bar{\mathcal{P} }};Q_{[N]}^{[\mathcal{P}]},A_{[N]}^{[\mathcal{P}]},\mathcal{P}) = 0 \label{Database Privacy}, \quad \forall \mathcal{P} 
\end{align}

An achievable MM-SPIR scheme is a scheme that satisfies the reliability constraint \eqref{Reliability}, the user privacy constraint \eqref{User Privacy} and the database privacy constraint \eqref{Database Privacy}. Similar to single-database SPIR \cite{SPIR}, single-database MM-SPIR is infeasible as well. In order to make single-database MM-SPIR feasible, we use the multi-message version of the extended SPIR formulation in \cite{SPIR_atPIR}, where the user is able to fetch a random subset of the shared server-side common randomness before the retrieval process starts, in the MM-SPIR setting.

\subsection{PSU} \label{PSU Formulation}
In the PSU problem, two parties each holding a dataset, wish to jointly compute the union of their sets without revealing anything else to either party. Let $\mathcal{A}$ denote the global alphabet. The first party $P_1$ stores a dataset $\Omega_1$ across its own $N_1 \geq 1$ replicated and non-colluding databases, and the second party $P_2$ stores a dataset $\Omega_2$ across its own $N_2 \geq 1$ replicated and non-colluding databases. Let $\mathcal{P}_1$ and $\mathcal{P}_2$ denote the realizations of the random variables $\Omega_1$ and $\Omega_2$, respectively. All elements in $\mathcal{P}_1$ and $\mathcal{P}_2$ are selected from $\mathcal{A}$ under an arbitrary statistical distribution, i.e., $\mathcal{P}_1, \mathcal{P}_2 \subseteq \mathcal{A}$. We denote one of the parties as the leader/server and the other as the client/user. Without loss of generality, let party $P_1$ be the server. Then, as in \cite{PSI_journal}, $P_1$ privacy, $P_2$ privacy and PSU reliability constraints jointly form a contradiction, and as in all SPIR formulations \cite{SPIR_ORI, SPIR}, the server databases need to share an amount of common randomness $\mathcal{R}_S$ besides their own datasets. Then, the party $P_2$ generates $N_1$ queries $Q^{[\mathcal{P}_2]}_{[N_1]}$ and sends them to the databases associated with the party $P_1$. After receiving the query $Q^{[\mathcal{P}_2]}_{n_1}$, the $n_1$th database of the party $P_1$ responds with an answer $A^{[\mathcal{P}_2]}_{n_1}$.

For each database in the party $P_1$, the answer $A^{[\mathcal{P}_2]}_{n_1}$ should be generated truthfully according to the received query, its own dataset and its own common randomness, 
\begin{align}
\text{[PSU deterministic answer]} \quad    H(A^{[\mathcal{P}_2]}_{n_1}|Q^{[\mathcal{P}_2]}_{n_1},\Omega_1,\mathcal{R}_S) = 0, \quad \forall n_1, ~ \forall \mathcal{P}_2
\end{align}

When the PSU process is complete, the party $P_2$ should be able to reliably compute the union $\Omega_1 \cup \Omega_2$ based on the sent queries, the collected answers and the knowledge of $\Omega_2$ without knowing $|\Omega_1 \cup \Omega_2|$ in advance. This is captured by the PSU reliability constraint,
\begin{align}\label{PSI Reliability}
\text{[PSU reliability]} \quad H(\Omega_1 \cup \Omega_2|Q^{[\mathcal{P}_2]}_{[N_1]},A^{[\mathcal{P}_2]}_{[N_1]},\Omega_2)=0, \quad \forall \mathcal{P}_2
\end{align}

The privacy requirements in PSU can be divided into two parts to protect each participating party: $P_1$ privacy and $P_2$ privacy. First, the party $P_2$ wants to protect $\Omega_1 \cup \Omega_2$, however, since the party $P_2$ does not know $\Omega_1$ when generating its queries, the queries cannot depend on $\Omega_1$, and thus, $P_2$ should only protect $\Omega_2$ in queries. Thus, the queries sent by $P_2$ should not leak any information about its own dataset, i.e., any individual database associated with $P_1$ learns nothing about $\Omega_2$ from all the information it has, 
\begin{align} \label{P2 privacy} 
\text{[PSU $P_2$ privacy]} \qquad I(\Omega_2;Q^{[\Omega_2]}_{n_1},A^{[\Omega_2]}_{n_1},\Omega_1,\mathcal{R}_S)=0, \quad \forall n_1
\end{align}

Because of the known and fixed global alphabet $\mathcal{A}$, it is obvious that we have the following two constraints $H(\Omega_2|\bar{\Omega}_2) = 0$ and $H(\bar{\Omega}_2|\Omega_2) = 0$, which lead to the following relationship,
\begin{align}
    H(\Omega_2) 
    &= H(\Omega_2) - H(\Omega_2|\bar{\Omega}_2) \\
    &= I(\Omega_2;\bar{\Omega}_2) \\
    &= H(\bar{\Omega}_2) - H(\bar{\Omega}_2|\Omega_2) \\
    &= H(\bar{\Omega}_2)
\end{align}
Thus, we obtain the following identity,
\begin{align}
    I(\bar{\Omega}_2&;Q^{[\Omega_2]}_{n_1},A^{[\Omega_2]}_{n_1},\Omega_1,\mathcal{R}_S) \nonumber \\
    &= H(\bar{\Omega}_2) - H(\bar{\Omega}_2|Q^{[\Omega_2]}_{n_1},A^{[\Omega_2]}_{n_1},\Omega_1,\mathcal{R}_S) \\
    &= H(\bar{\Omega}_2) - H(\bar{\Omega}_2|Q^{[\Omega_2]}_{n_1},A^{[\Omega_2]}_{n_1},\Omega_1,\mathcal{R}_S) + H(\bar{\Omega}_2|Q^{[\Omega_2]}_{n_1},A^{[\Omega_2]}_{n_1},\Omega_1,\Omega_2,\mathcal{R}_S) \\
    &= H(\bar{\Omega}_2) - I(\bar{\Omega}_2;\Omega_2|Q^{[\Omega_2]}_{n_1},A^{[\Omega_2]}_{n_1},\Omega_1,\mathcal{R}_S) \\
    &= H(\Omega_2) - H(\Omega_2|Q^{[\Omega_2]}_{n_1},A^{[\Omega_2]}_{n_1},\Omega_1,\mathcal{R}_S) + H(\Omega_2|Q^{[\Omega_2]}_{n_1},A^{[\Omega_2]}_{n_1},\Omega_1,\bar{\Omega}_2,\mathcal{R}_S) \\
    &= H(\Omega_2) - H(\Omega_2|Q^{[\Omega_2]}_{n_1},A^{[\Omega_2]}_{n_1},\Omega_1,\mathcal{R}_S) \\
    &= I(\Omega_2;Q^{[\Omega_2]}_{n_1},A^{[\Omega_2]}_{n_1},\Omega_1,\mathcal{R}_S)
\end{align}
As a consequence, we obtain the following equivalent expression for $P_2$ privacy,
\begin{align} \label{P2 privacy2} 
\text{[PSU $P_2$ privacy]} \qquad I(\bar{\Omega}_2;Q^{[\Omega_2]}_{n_1},A^{[\Omega_2]}_{n_1},\Omega_1,\mathcal{R}_S) = 0, \quad \forall n_1
\end{align}

From the union result $\Omega_1 \cup \Omega_2$, the party $P_2$ always knows that the party $P_1$ contains the elements in $(\Omega_1 \cup \Omega_2) \backslash \Omega_2$ and does not contain the elements in $\overline{(\Omega_1 \cup \Omega_2)}$. Noting that $(\Omega_1 \cup \Omega_2) \backslash \Omega_2 \cup \overline{(\Omega_1 \cup \Omega_2)}  = \bar{\Omega}_2$, thus, $P_2$ should learn nothing about whether $P_1$ contains any element in $\Omega_2$ (we denote this information by $E_{1,\Omega_2}$) from the generated queries, the collected answers and its own dataset,
\begin{align}\label{P1 privacy} 
\text{[PSU $P_1$ privacy]} \quad  I(E_{1,\Omega_2};Q^{[\mathcal{P}_2]}_{[N_1]},A^{[\mathcal{P}_2]}_{[N_1]},\Omega_2)=0, \quad \forall \mathcal{P}_2
\end{align}

\begin{theorem}\label{thm1}
PSU is equivalent to MM-SPIR with $L = 1$ and $P = |\mathcal{A}|- |\Omega_2|$.
\end{theorem}

\begin{Proof}
We prove the equivalence between PSU and MM-SPIR similar to the proof of equivalence between PSI and MM-SPIR in \cite{PSI_journal} after mapping the dataset in each party to a corresponding incidence vector. Specifically, the $P_1$ privacy, $P_2$ privacy and PSU reliability constraints in the PSU problem are consistent with the database privacy, user privacy and reliability constraints in the MM-SPIR problem if $\bar{\Omega}_2$ in PSU is treated as $\Omega$ in MM-SPIR. By contrast, the consistency of the three constraints of PSI and MM-SPIR is true if $\Omega_2$ in PSI is treated as $\Omega$ in MM-SPIR.
\end{Proof}

\begin{remark}
From \cite{PSI_journal}, we note that PSI is equivalent to MM-SPIR with with $L = 1$ and $P =|\Omega_2|$. From Theorem~\ref{thm1} above, we note that PSU is equivalent to MM-SPIR with $L = 1$ and $P = |\mathcal{A}|- |\Omega_2|$.\footnote{These two conclusion are built upon the assumption that the party $P_2$ is the user. As an alternative, if the party $P_1$ is treated as the user, we just need to replace $\Omega_1$ with $\Omega_2$ in these two statements.} From the de Morgan's law, which says $\overline{A \cup B} = \overline{A} \cap \overline{B}$, we have that $A \cup B = \overline{\overline{A} \cap \overline{B}}$, thus, the set union can be obtained by a composition of set intersection and set complement. This shows the duality between PSU and PSI problems. We note, however, that the parties should agree on whether they will perform PSU or PSI, as the specific protocol will depend on it. In this paper, we focus on designing specific PSU protocols.
\end{remark}

\begin{remark}
In certain applications of PSU, one or both of the parties may have only a single database. Since PSU is equivalent to MM-SPIR from Theorem~\ref{thm1}, and since single-database MM-SPIR is infeasible \cite{SPIR}, in such cases, one of the two parties may obtain (fetch) a random subset of the shared server-side common randomness from the other party prior to the start of the PSU process, as in \cite{SPIR_atPIR}. This makes MM-SPIR feasible, and thus, PSU feasible.
\end{remark}

\begin{remark}
As PSI was generalized to multi-party PSI (MP-PSI) in \cite{MP-PSI_journal}, PSU can be generalized to MP-PSU. As in MP-PSI, MP-PSU will require additional common randomness allocation among the clients. To avoid repetition, we skip the detailed development of MP-PSU, however, in the next subsection, we present a particular MP-PSU in detail, where one party has no input. As a critical difference, in the reliability verification stage, we need to have the sum in \cite[Eqn.~(56)]{MP-PSI_journal} equal to 0 if all the clients contain the same element in the MP-PSI problem while this sum should be 0 if none of the clients contain this element in the MP-PSU problem; see Example~\ref{MP-PSU1} for details. In MP-PSU, if all the parties have a single database, we can construct an achievable scheme by using pre-fetched server-side common randomness from the leader party as in \cite{SPIR_atPIR}. In addition, for common randomness allocation among the clients, we make use of the distributed property of non-colluding databases as well as the RSPIR approach introduced in \cite{RSPIR}.
\end{remark}

\subsection{Private Distributed FSL} \label{FSL Formulation}
\begin{figure}[t]
\centering
\includegraphics[width=0.82\linewidth]{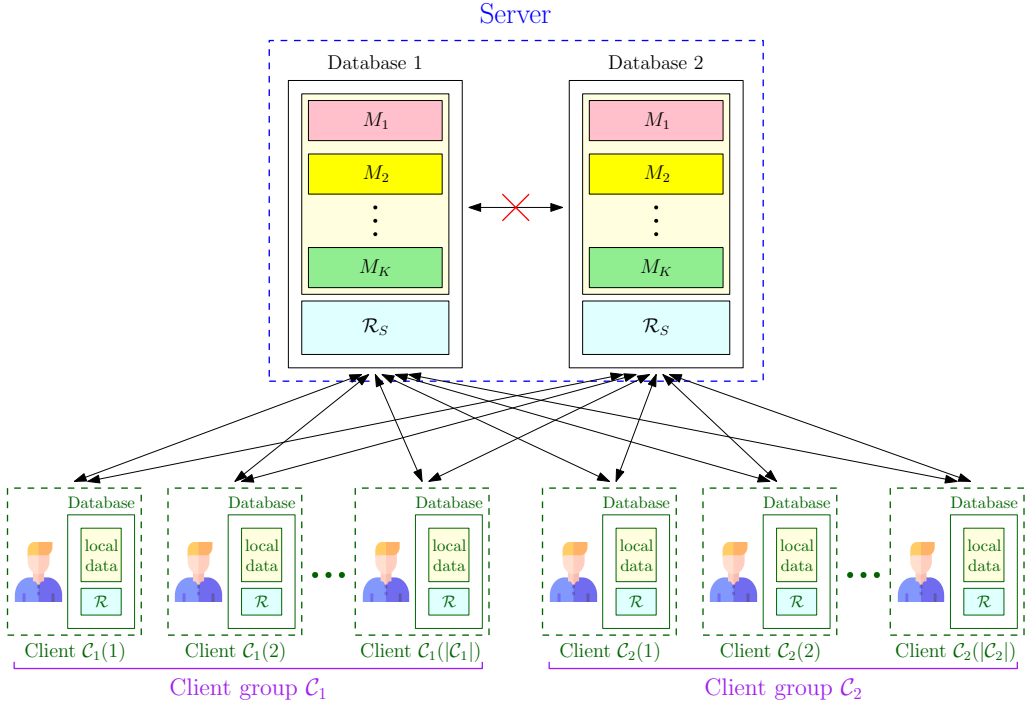}
\caption{Distributed federated submodel learning (FSL) system model.}
\label{System model}
\end{figure}

We consider a distributed FSL problem with one server that contains $N=2$ non-colluding and replicated databases\footnote{We start this investigation with the simplest case of two databases. Our achievable scheme works for any number of databases after minor modifications. However, how to improve the performance by increasing the number of databases needs further study.}, and $C$ clients that are selected by the server to participate in one round of the FSL process; see Fig.~\ref{System model}. By convention, every client establishes a direct secure and authenticated communication channel with both databases. The full model for learning stored at the server side comprises $K$ submodels, each one of which consisting of $L$ i.i.d.~symbols that are uniformly selected from a finite field $\mathbb{F}_q$. Thus, each database in the server contains the full model $M_{[K]}$, and we have,
\begin{align}
    H(M_k) &= L, \quad \forall k \label{Submodel Length} \\
    H(M_{[K]}) &= H(M_1) + \dots + H(M_K)  = KL \label{Submodel IID}
\end{align}

The two databases also share an amount of server-side common randomness $\mathcal{R}_S$ that is unknown to the clients. Each selected client is interested in updating one or more submodels according to its local training data. Specifically, for $i \in [C]$, the $i$th client wishes to update the submodels whose index set is denoted by the random variable $\Gamma^{\langle i \rangle}$, whose realization is denoted by $\gamma^{\langle i \rangle}$. For $i \in [C]$, the random variable $Y^{\langle i \rangle} = \{Y^{\langle i \rangle}_1,Y^{\langle i \rangle}_2,\dots,Y^{\langle i \rangle}_K\}$ is used to denote the corresponding incidence vector of $\Gamma^{\langle i \rangle}$ after mapping to the alphabet as in \cite{PSI_journal,MP-PSI_journal}. 

\begin{figure}[t]
\centering
\begin{minipage}{.45\linewidth}
  \includegraphics[height=8.25em]{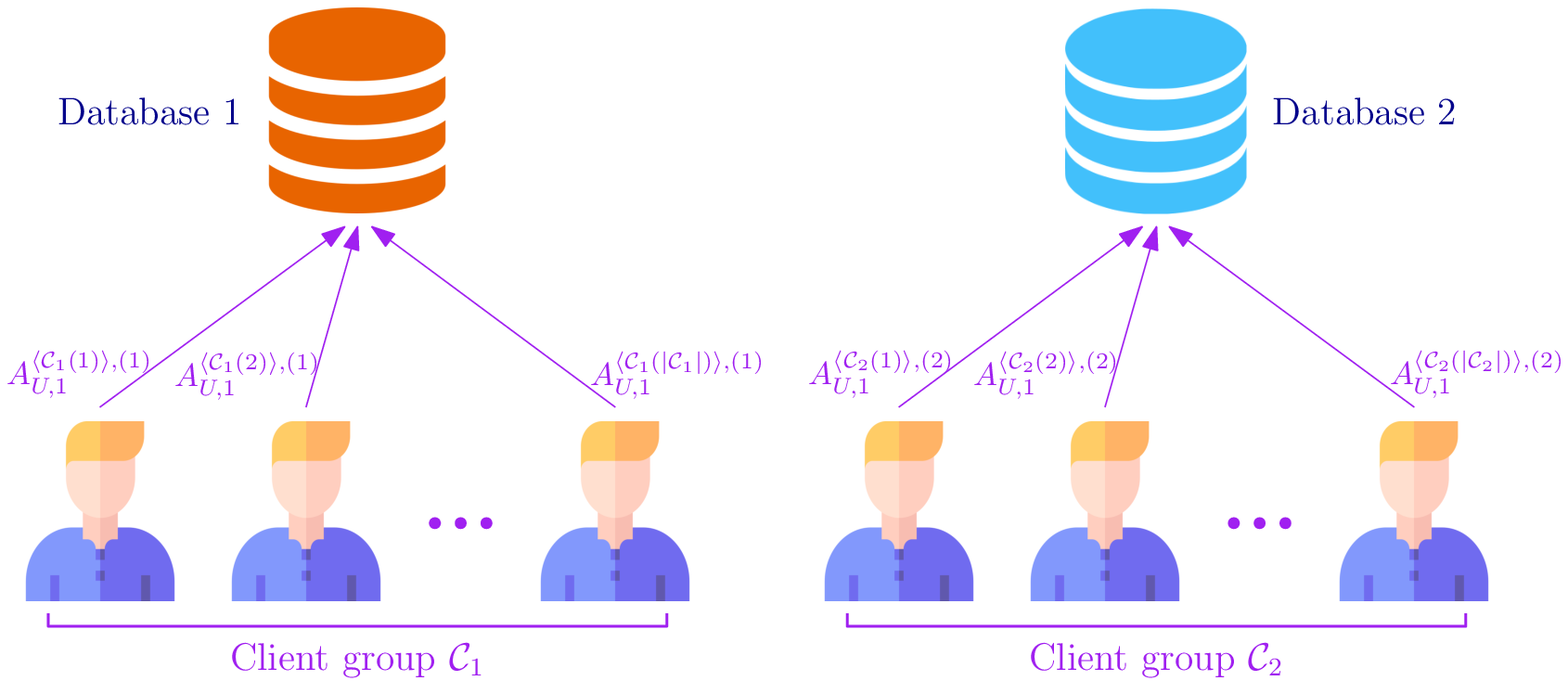}
  \caption*{(a) FSL-PSU phase step~1.}
\end{minipage}
\hspace{.05\linewidth}
\begin{minipage}{.45\linewidth}
  \includegraphics[height=8.25em]{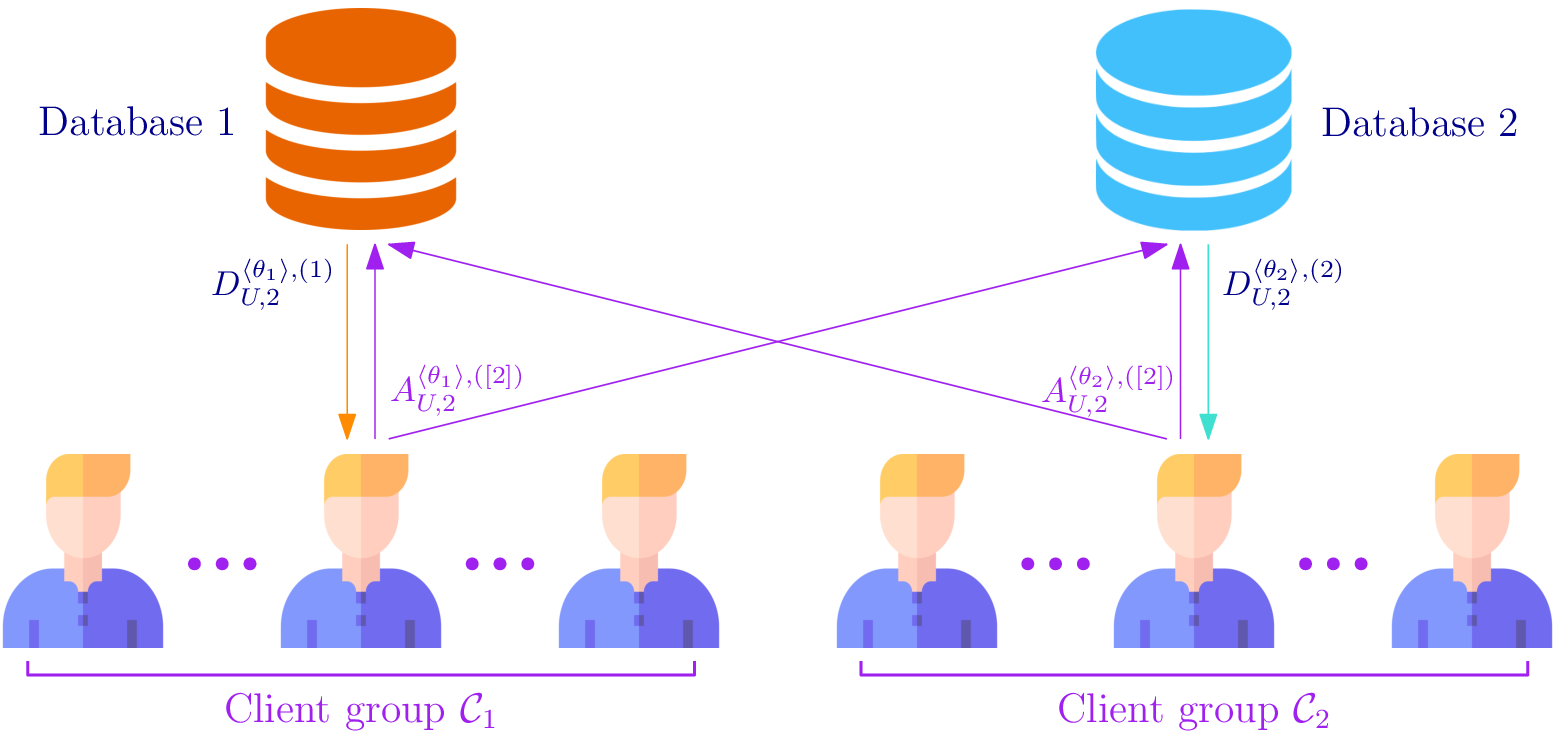}
  \caption*{(b) FSL-PSU phase step~2.}
\end{minipage}
\caption{Data flow in the FSL-PSU phase of our FSL system model.}
\label{FSL_PSU_fig}
\end{figure}

We formulate our FSL process following the seminal FSL work in \cite{FSL}. At the beginning, each individual database in the server needs to calculate the union of the selected clients' desired submodel index sets $\Gamma^{\langle 1 \rangle} \cup \Gamma^{\langle 2 \rangle} \cup \cdots \cup \Gamma^{\langle C \rangle}$ denoted by $\Gamma$. This phase is referred to as the FSL-PSU phase. Due to the constraint that the two databases in the server cannot communicate with each other directly, our solution is to use randomly selected alive clients as intermediators to route the information received by the two databases rather than to enforce each client to send the same answer to both databases. The main objective of this new approach is to reduce the total communication cost and the needed communication time. Thus, we separate $C$ clients into two groups: a group of clients whose index set denoted by $\mathcal{C}_1 = \{\mathcal{C}_1(1), \mathcal{C}_1(2), \dots, \mathcal{C}_1(|\mathcal{C}_1|)\}$ are associated with database $1$, and the other group of clients whose index set denoted by $\mathcal{C}_2 = \{\mathcal{C}_2(1),\mathcal{C}_2(2),\dots,\mathcal{C}_2(|\mathcal{C}_2|)\}$ are associated with database $2$. A potential separation method is to rely on each client's communication channel bandwidth (or quality) with the two databases. For instance, a client is classified as belonging to $\mathcal{C}_1$ if its channel with database $1$ has a higher bandwidth (quality) than the channel with database $2$. Otherwise, this client is considered as belonging to $\mathcal{C}_2$. Note that $\mathcal{C}_1 \cap \mathcal{C}_2 = \emptyset$ and $\mathcal{C}_1 \cup \mathcal{C}_2 = [C]$. Please see Figs.~\ref{System model} and \ref{FSL_PSU_fig} for depictions.

The FSL-PSU phase is further divided into two steps considering the fact that two random clients (one from each client group) are utilized to relay the information between the databases; see Fig.~\ref{FSL_PSU_fig}. This information is produced from the answers that are collected by the two databases individually from their associated client groups. In the first step, there is no need for each client to download any information from the databases since the server itself is not involved in the PSU computation, namely the downloads $D^{\langle \mathcal{C}_j \rangle, (j)}_{U,1}$ are null\footnote{In this work, we use the value in $\langle \rangle$ to denote the index of client and the value in $()$ to denote the index of database for clarity. The superscript of the download $D$ or  the answer $A$ in the following text implies the information flow during the client-database communication. The first subscript of $D$ or $A$ is used to show it is within the FSL-PSU phase or FSL-write phase (the letter U stands for union and the letter W stands for write), whereas the second subscript is used to denote the step number within this phase. In particular, ``$D^{\langle \mathcal{C}_j \rangle,(j)}_{U,1}$ are null'' here means that the communication between any client in $\mathcal{C}_j$ and database $j$ is always empty in the first step of FSL-PSU phase.} for all $j \in  [2]$. As a consequence, the only operation in this step is to make clients send their well-designed answers $A^{\langle \mathcal{C}_j \rangle,(j)}_{U,1}$ to the associated database. In the second step, for all $j \in [2]$, database $j$ processes the answers received from its associated clients with the aid of its own server-side common randomness, and then the produced $D^{\langle \theta_j \rangle,(j)}_{U,2}$ is merely downloaded by a randomly chosen client whose index is $\theta_j$ within its associated client group $\mathcal{C}_j$. Finally, client $\theta_j$ forwards the same processed answer $A^{\langle \theta_j \rangle,([2])}_{U,2}$ based on the received download to both databases; see Fig.~\ref{FSL_PSU_fig}.  

Similar to the conventional multi-user PIR/SPIR problem formulated in \cite{DoubleBlind_PIR,MultiBlind_SPIR}, the constraints accompanying FSL-PSU phase comprises three parts. First, each database $j$ should be able to reliably determine the union $\Gamma$ using all the collected answers $\{A^{\langle \mathcal{C}_j \rangle,(j)}_{U,1},A^{\langle \theta_{[2]} \rangle,(j)}_{U,2}\}$ within two FSL-PSU steps and its own server-side common randomness $\mathcal{R}_S$, which is captured by,
\begin{align} \label{FSL-PSU reliability}
    \text{[FSL-PSU reliability]} \quad H(\Gamma|A^{\langle \mathcal{C}_j \rangle,(j)}_{U,1},A^{\langle \theta_{[2]} \rangle,(j)}_{U,2},\mathcal{R}_S)=0, \quad \forall j
\end{align}

Second, the databases should not learn anything further about the set $\Gamma^{\langle [C] \rangle}$ or $Y^{\langle [C] \rangle}$ other than the union $\Gamma$. Note that if an element is not in the union $\Gamma$, each database concludes that no client contains this element. Otherwise, each database learns that at least one client contains this element. Let $Y^{\langle i \rangle}_{\Gamma} = \{Y^{\langle i \rangle}_k\!\!: k \in \Gamma\}$, we define a new set $Y_{\Gamma} = Y^{\langle [C] \rangle}_{\Gamma}$, then,
\begin{align} \label{FSL-PSU privacy}
    \text{[FSL-PSU privacy]} \quad I(Y_{\Gamma};A^{\langle \mathcal{C}_j \rangle,(j)}_{U,1},A^{\langle \theta_{[2]} \rangle,(j)}_{U,2},\mathcal{R}_S|\sum_{i \in [C]} Y^{\langle i \rangle}_k > 0, \forall k \in \Gamma ) = 0, \quad \forall j
\end{align}

Third, client $\theta_j$ that obtains the download $D^{\langle \theta_j \rangle,(j)}_{U,2}$ from database $j$ should learn nothing about the other clients’ desired submodel indices. Hence, we have the following constraint,
\begin{align} \label{FSL-PSU inter-client privacy}
    \text{[FSL-PSU inter-client privacy]} \quad I(Y^{\langle [C] \backslash \theta_j \rangle}_{\Gamma};D^{\langle \theta_j \rangle,(j)}_{U,2},Y^{\langle \theta_j \rangle}) = 0, \quad \forall j
\end{align}

A valid FSL-PSU phase is one that satisfies the FSL-PSU reliability \eqref{FSL-PSU reliability}, the FSL-PSU privacy \eqref{FSL-PSU privacy} and the FSL-PSU inter-client privacy \eqref{FSL-PSU inter-client privacy}. The efficiency of an FSL-PSU phase is measured in terms of the number of bits in the involved communication strings. Therefore, for the FSL-PSU phase itself, we wish to reduce the total number of bits in the answers $\{A^{\langle \mathcal{C}_1 \rangle,(1)}_{U,1},A^{\langle \mathcal{C}_2 \rangle,(2)}_{U,1},A^{\langle \theta_1 \rangle,([2])}_{U,2},A^{\langle \theta_2 \rangle,([2])}_{U,2}\}$ and downloads $\{D^{\langle \theta_1 \rangle,(1)}_{U,2},D^{\langle \theta_2 \rangle,(2)}_{U,2}\}$ to the extent possible.

\begin{figure}[t]
\centering
\begin{minipage}{.45\linewidth}
  \includegraphics[height=8.4em]{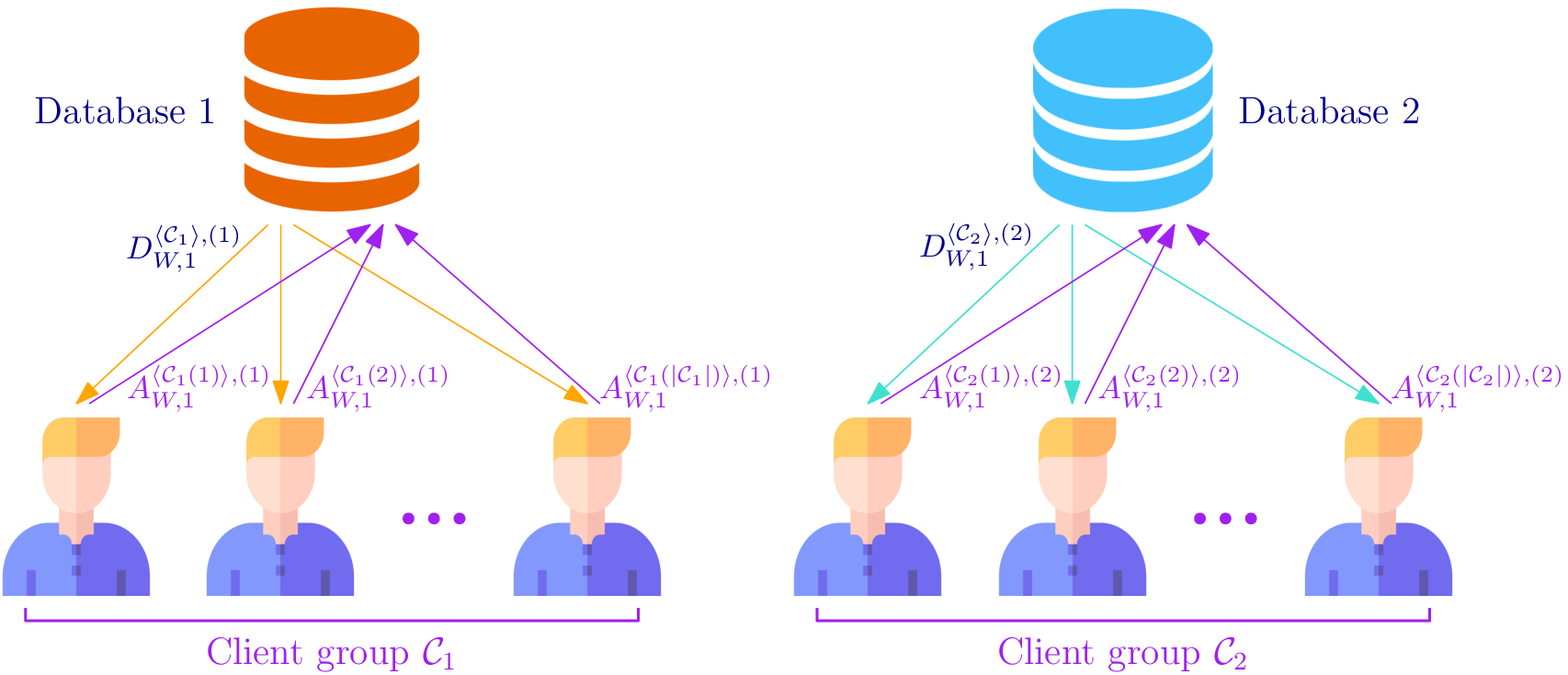}
  \caption*{(a) FSL-write phase step 1.}
\end{minipage}
\hspace{.05\linewidth}
\begin{minipage}{.45\linewidth}
  \includegraphics[height=8.4em]{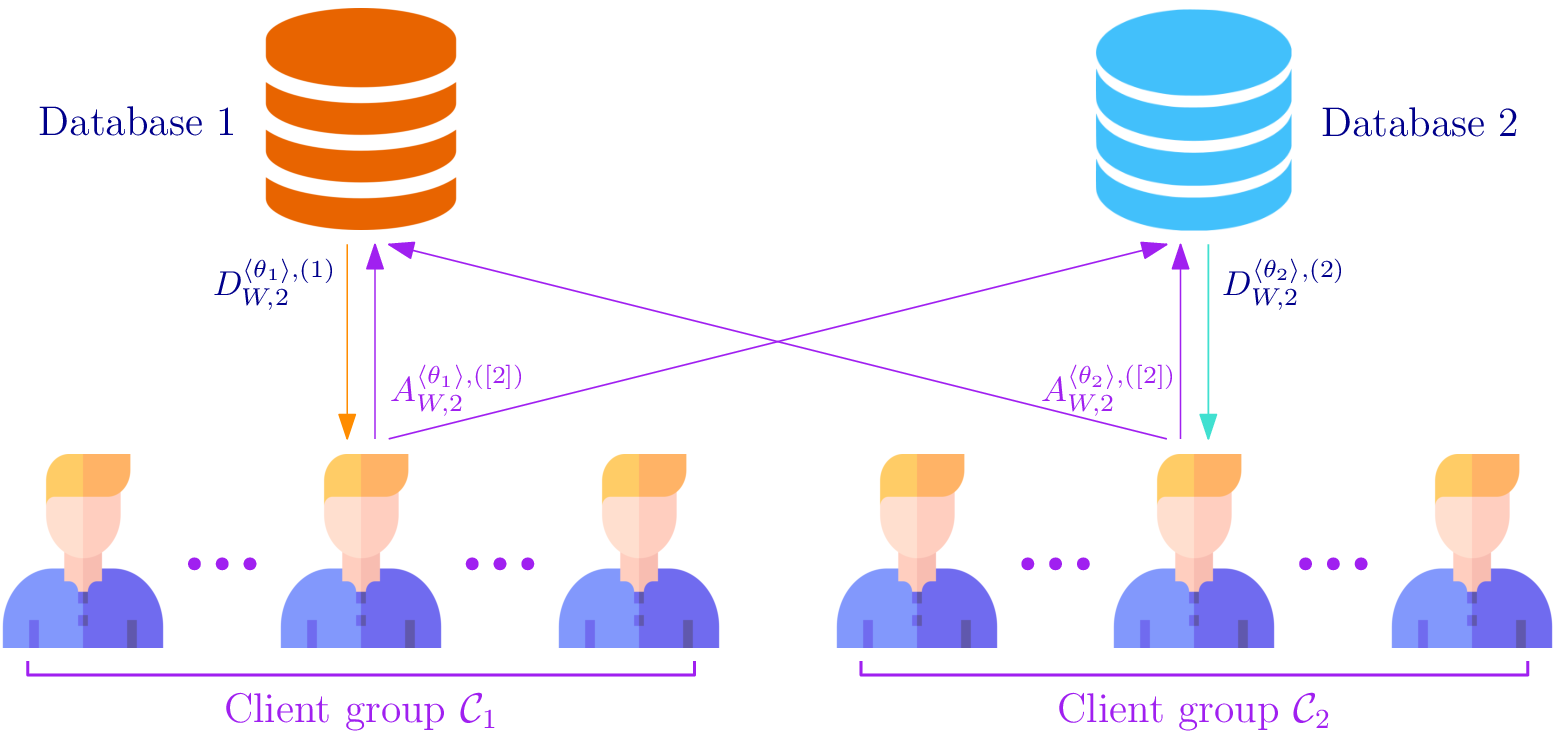}
  \caption*{(b) FSL-write phase step 2.}
\end{minipage}
\caption{Data flow in the FSL-write phase of our FSL system model.}
\label{FSL_Write_fig}
\end{figure}

When the FSL-PSU phase is completed, each database will learn $\Gamma$, the union of the submodel indices to be updated. Next, we proceed to the FSL-write phase where each database will update the full learning model synchronously. The FSL-write phase is analogous to the FSL-PSU phase, and therefore, is also divided into two steps as the FSL-PSU phase. The difference is that in the first step, both databases broadcast the same set of submodels $M_{\Gamma} = \{M_k\!\!: k \in \Gamma\}$ to their associated clients before each client trains its desired submodel set $M_{\Gamma^{\langle i \rangle}}$ by employing its local data. Hence, for all $j \in [2]$, the downloads $D^{\langle \mathcal{C}_j \rangle,(j)}_{W,1}$ are always in the form of $M_{\Gamma}$. Subsequently, clients send their well-processed answer $A^{\langle \mathcal{C}_j \rangle,(j)}_{W,1}$ corresponding to the submodel updates back to the associated database. In the second step, for all $j \in [2]$, database $j$ processes its associated clients' answers through different server-side common randomness and then the produced $D^{\langle \theta_j \rangle,(j)}_{W,2}$ is downloaded by the $\theta_j$th client again. Finally, the $\theta_j$th client forwards the same resulting answer $A^{\langle \theta_j \rangle,([2])}_{W,2}$ to both databases after processing the newly received download; see Fig.~\ref{FSL_Write_fig}.

Likewise, the constraints accompanying FSL-write phase comprises three parts. First, each database $j$ should be able to reliably obtain the aggregation of all the submodel updates according to all the collected answers $\{A^{\langle \mathcal{C}_j \rangle,(j)}_{W,1},A^{\langle \theta_{[2]} \rangle,(j)}_{W,2}\}$ within two FSL-write steps, its own current full model $M_{[K]}$ and its own server-side common randomness $\mathcal{R}_S$. When the submodel training by means of the local data in the $i$th client is complete, for all $k \in \Gamma^{\langle i \rangle}$, this client will generate the increment $\Delta_k = \{\Delta^{\langle i \rangle}_{k,1},\Delta^{\langle i \rangle}_{k,2},\dots,\Delta^{\langle i \rangle}_{k,L}\}$ for each symbol in the submodel $M_k = \{M_{k,1}, M_{k,2},\dots,M_{k,L}\}$. Thus, for the $k$th submodel, let $\Phi_k$ be the set of clients whose desired submodel index set $\Gamma^{\langle i \rangle}$ contains $k$, its correct updated version should be $M^{\prime}_k = \{M^{\prime}_{k,1},M^{\prime}_{k,2},\dots,M^{\prime}_{k,L}\} = \{M_{k,1}+\sum_{i \in \Phi_k} \Delta^{\langle i \rangle}_{k,1},M_{k,2}+\sum_{i \in \Phi_k} \Delta^{\langle i \rangle}_{k,2},\dots,M_{k,L}+\sum_{i \in \Phi_k} \Delta^{\langle i \rangle}_{k,L}\}$. For each database in the server, the correct updated submodel aggregation should be $M^{\prime}_{\Gamma} = \{M^{\prime}_k\!\!: k \in \Gamma$\} and thus the first constraint can be expressed as,
\begin{align} \label{FSL-write reliability}
    \text{[FSL-write reliability]} \quad H(M^{\prime}_{\Gamma}|A^{\langle \mathcal{C}_j \rangle,(j)}_{W,1},A^{\langle \theta_{[2]} \rangle,(j)}_{W,2},M_{[K]},\mathcal{R}_S)=0, \quad \forall j
\end{align}

Second, no database should learn any knowledge about each client's desired submodel index set or any further information beyond the updated submodel aggregation about each client's submodel increment. For the $i$th client's submodel increment, let $\Delta^{\langle i \rangle}_{\Gamma} = \{\Delta^{\langle i \rangle}_{k,l} \!\!: k \in \Gamma, l \in [L]\}$, we define a new set $\Delta_{\Gamma} = \Delta^{\langle [C] \rangle}_{\Gamma}$, then,\footnote{In general, the first term in the following conditional mutual information should be $Y_{\Gamma},\Delta_{\Gamma}$ rather than $\Delta_{\Gamma}$. In the FSL-write phase, we note that the information transmission only involves the submodel increment regarding $M_{\Gamma}$ and it has nothing to do with the incidence vectors $Y^{\langle [C]\rangle}$. That means that if a database learns nothing beyond the aggregation increment from all the selected clients, this database definitely learns nothing about the incidence vector $Y_{\Gamma}$. Therefore, the expression $\Delta_{\Gamma}$ takes the place of $Y_{\Gamma},\Delta_{\Gamma}$. This observation also applies to the FSL-write inter-client privacy constraint \eqref{FSL-write inter-client privacy} in which the expression $\Delta^{\langle [C] \backslash \theta_j \rangle}_{\Gamma}$ is used in place of $Y^{\langle [C] \backslash \theta_j \rangle}_{\Gamma},\Delta^{\langle [C] \backslash \theta_j \rangle}_{\Gamma}$.}
\begin{align} \label{FSL-write privacy}
    \!\text{[FSL-write privacy]} \ I(\Delta_{\Gamma};A^{\langle \mathcal{C}_j \rangle,(j)}_{W,1},A^{\langle \theta_{[2]} \rangle,(j)}_{W,2},M_{[K]},\mathcal{R}_S|\sum_{i \in \Phi_k} \!\! \Delta^{\langle i \rangle}_{k,l}, \forall k \! \in \! \Gamma, \forall l \! \in \! [L]) = 0, \ \forall j
\end{align}

Third, the $\theta$th client should learn nothing about the other clients’ desired submodel indices or submodel increments according to its obtained download $D^{\langle \theta_j \rangle,(j)}_{W,2}$ from database $j$. Hence, we have the following constraint,
\begin{align} \label{FSL-write inter-client privacy}
     \text{[FSL-write inter-client privacy]} \quad I(\Delta^{\langle [C] \backslash \theta_j \rangle}_{\Gamma};D^{\langle \theta_j \rangle,(j)}_{W,2},\Gamma^{\langle \theta_j \rangle},\Delta_{\Gamma^{\langle \theta_j \rangle}}) = 0
\end{align}

A valid FSL-write phase is a one that satisfies the FSL-write reliability \eqref{FSL-write reliability}, the FSL-write privacy \eqref{FSL-write privacy} and the FSL-write inter-client privacy \eqref{FSL-write inter-client privacy}. Given any specific FSL problem with fixed initial parameters, the communication cost of sending the set of submodels $M_{\Gamma}$ to each client from the two databases is a constant. Hence, the efficiency of an FSL-write phase is also measured in terms of the total number of bits in the answers $\{A^{\langle \mathcal{C}_1 \rangle,(1)}_{W,1},A^{\langle \mathcal{C}_2 \rangle,(2)}_{W,1},A^{\langle \theta_1 \rangle,([2])}_{W,2},A^{\langle \theta_2 \rangle,([2])}_{W,2}\}$ and downloads $\{D^{\langle \theta_1 \rangle,(1)}_{W,2},D^{\langle \theta_2 \rangle,(2)}_{W,2}\}$, and we wish to minimize it as much as possible. If we do not consider the generation of client-side common randomness that is necessary to perform the FSL, one complete FSL round consists of two phases: FSL-PSU phase and FSL-write phase. Our objective is to make the total number of communication bits exchanged in these two phases as small as possible. Further, this FSL round can be executed in an iterative manner until a predefined termination criterion is satisfied, e.g., the accuracy of the updated global model exceeds the preset threshold or a preset maximal number of iterations is reached.

\section{Main Result}
Our main result is a new private FSL algorithm as described above. The following theorem gives its performance in terms of the total communication cost in the entire process including the cost of FSL-PSU, FSL-write, and the generation of the necessary common randomness at the clients. The proof of the theorem is given in Section~\ref{FSL-PSU Phase} and Section~\ref{FSL-write Phase}.

\begin{theorem}
The total communication cost of the proposed private FSL scheme in one round is $\mathcal{O}(CK+C|\Gamma|L)$ in $q$-ary bits, where $C$ is the number of selected clients, $K$ is the total number of submodels, and $|\Gamma|$ is the number of updated submodels in the given round. Here, $\mathcal{O}(CK)$ is due to the FSL-PSU phase, while $\mathcal{O}(C|\Gamma|L)$ is due to the FSL-write phase.
\end{theorem}

\begin{remark}
The achievability of the theorem starts with an MM-SPIR with multiple replicated and non-colluding databases. The storage in the databases is uncoded and without noise. We unify PSU and secure aggregation in a common information theoretic framework, and propose a novel private FSL scheme. We take advantage of the non-colluding aspect of the databases to implement simple common randomness generation/distribution across selected clients. 
\end{remark}

\begin{remark}
Our proposed FSL achieves unconditional information theoretic privacy. This is different from most prior secure aggregation works that focus on the computational security, e.g., \cite{SecAgg, FastSecAgg, PMG_Agg, Prio, FSL}. It is also different from prior private read update write (PRUW) works \cite{Kim_FSL, XSTFSL, Sajani_FSL1, Sajani_FSL2, Sajani_FSL3, Sajani_FSL4} in which only a single client at a time updates the full model in an FSL round, although information theoretic security is satisfied. Our proposed private FSL scheme is robust against client drop-outs, client late-arrivals, and database drop-outs. Moreover, there is no constraint on the number of clients that may drop out during the FSL process.
\end{remark}

\begin{remark}
The communication cost of our proposed private FSL, $\mathcal{O}(CK+C|\Gamma|L)$, outperforms the best-known communication cost in the existing literature \cite{SecAgg,SecAgg+,TurboAgg,FastSecAgg}, which is at least $\mathcal{O}(CKL)$. In the seminal FSL work \cite{FSL}, the communication cost is $\mathcal{O}(C|\Gamma|)$ for the PSU phase and $\mathcal{O}(C|\Gamma|L)$ for the whole FSL process with much weaker privacy guarantee. Although this communication cost is a little better than our communication cost in terms of the PSU phase, the PSU \cite{FSL} yields erroneous results while our PSU yields completely accurate results. Furthermore, the PSU problem and the subsequent secure aggregation problem are considered separately in \cite{FSL}. We note that the total number of submodels $K$ is very large when each product is represented by an individual submodel in the e-commerce recommendation system in \cite{FSL}. Thus, given the scale of the full learning model and the general average size of clients' desired products in practice, we can further optimize the communication cost by adjusting the size of $K$, e.g., combining relevant products into the same goods category. Specifically, as we decrease $K$, the product of $|\Gamma|$ and $L$ will likely increase such that $K$ and $|\Gamma|L$ will have the same order. Thus, the communication cost of our scheme is superior to existing schemes, and can be further improved by optimizing the system model parameters. However, it is difficult to find a fair metric to compare our communication cost with the ones in \cite{XSTFSL,Sajani_FSL_Trans}. The main reason for this is that the schemes in \cite{XSTFSL,Sajani_FSL_Trans} require that only one client updates one submodel at a time, and also heavily rely on the sufficiently large number of databases $N$. That is, the schemes in \cite{XSTFSL,Sajani_FSL_Trans} require at least $N\geq 4$ databases, and cannot be compared to the scheme in our paper where the number of databases is $N=2$. If we follow the asymptotic assumption $L \gg K$ and let $C$ take value $1$, the only conclusion we can draw is that, the communication cost in these two different schemes are both a linear function of the submodel size $L$.
\end{remark}

\begin{remark}
Generally, the existing private FL schemes in the computer science literature rely on heavy cryptographic computations, while our proposed FSL scheme relies only on simple addition and multiplication computations in the finite field $\mathbb{F}_q$, at both client and server sides. In addition, due to unstable inter-client communications in practice, and the impermissible inter-database communication in our assumption, our FSL scheme relies only on client-database communications. In order to alleviate the challenges arising from the lack of inter-database communications, as a novel approach in our FSL scheme, we utilize random clients to route the required information between the databases in the server. The routed information comes from the answers collected by each database from its associated clients, and we further protect this information between the clients (inter-client privacy).
\end{remark}

\begin{remark}
In practical implementations, for each client, the upload speeds are typically much slower than download speeds during the client-database communications. Unlike the classical secure aggregation scheme in \cite{SecAgg}, the total communication time in our FSL process is further improved, since almost all of the alive clients send only one answer to one database in each phase. In addition, while determining the two client groups to be connected to the two databases, we can further improve the total communication time based on the actual bandwidth/quality of each client-database communication channel.
\end{remark}

\begin{remark}
The proposed private FSL scheme can be used iteratively in multiple rounds of an FSL process by refreshing server-side and client-side common randomness.
\end{remark}

\section{Examples for Blocks of Private Distributed FSL} \label{Example Section}
In this section, we give examples to explain the functionalities of the modules (boxes) in Fig.~\ref{Roadmap}. The examples get progressively more complex: Example~\ref{two-party PSU1} considers a two-party PSU setting where the client party has multiple databases and the leader party has a single database. Example~\ref{two-party PSU2} considers the slightly more difficult version of Example~\ref{two-party PSU1}, in that the client party also has a single database. In this case, single-database SPIR is infeasible, and the leader party needs to fetch client-side common randomness to use as leader-side side information as in \cite{SPIR_atPIR}. Example~\ref{MP-PSU1} considers generalized version of Example~\ref{two-party PSU2} to a multi-party (MP) case; in particular, there are 5 parties and each party has a single database. Example~\ref{MP-PSU2} is an extended version of Example~\ref{MP-PSU1}, where the leader party has two databases. This example reflects how the FSL-PSU phase of the proposed private FSL scheme works. Finally, Example~\ref{FSL} shows how the private write works. The FSL-PSU in Example~\ref{MP-PSU2} and the FSL-write in Example~\ref{FSL} together constitute our proposed private FSL scheme. 

\begin{example} \label{two-party PSU1}
\textbf{Two-party PSU; two-database client; one-database leader:} Consider a two-party PSU problem with a global alphabet $\mathcal{A} = \{1,2,3,4\}$. The first party $P_1$ contains  element $1$ and element $2$, i.e., $\mathcal{P}_1 = \{1,2\}$. The second party $P_2$ contains  element $1$ and element $3$, i.e., $\mathcal{P}_2 = \{1,3\}$. For convenience, the total number of elements in each party is public knowledge. The parties want to jointly compute the union of their element sets without revealing anything else to each other. By mapping their element sets into the corresponding incidence vectors, two parties construct the vectors as follows,
\begin{align}
    &\mbox{Party} ~ P_1: \quad \mathcal{P}_1 = \{1,2\} \quad \Rightarrow \quad X^{\langle 1 \rangle} = [X^{\langle 1 \rangle}_1 \:\: X^{\langle 1 \rangle}_2 \:\: X^{\langle 1 \rangle}_3 \:\: X^{\langle 1 \rangle}_4]^T = [1 \:\: 1 \:\: 0 \:\: 0]^T \\
    &\mbox{Party} ~ P_2: \quad \mathcal{P}_2 = \{1,3\} \quad \Rightarrow \quad X^{\langle 2 \rangle} = [X^{\langle 2 \rangle}_1 \:\: X^{\langle 2 \rangle}_2 \:\: X^{\langle 2 \rangle}_3 \:\: X^{\langle 2 \rangle}_4]^T = [1 \:\: 0 \:\: 1 \:\: 0]^T
\end{align}

First, $P_2$ (leader party) asks for the value of $X^{\langle 1 \rangle}_2$ from $P_1$ using an SPIR approach. $P_1$ (client party) has two replicated and non-colluding databases. These two databases share a server-side common randomness symbol $S_1$ that is uniformly selected from the finite field $\mathbb{F}_2$ and unknown to $P_2$. As a consequence, $P_1$ generates the answer table for two individual databases in the following form,
\begin{align}
    &A^{(1)}_U(1) = S_1,  &&A^{(2)}_U(1) = X^{\langle 1 \rangle}_1 + S_1 \\
    &A^{(1)}_U(2) = X^{\langle 1 \rangle}_1+X^{\langle 1 \rangle}_2+S_1,  &&A^{(2)}_U(2) = X^{\langle 1 \rangle}_2+S_1 \\
    &A^{(1)}_U(3) = X^{\langle 1 \rangle}_1+X^{\langle 1 \rangle}_3+S_1,  &&A^{(2)}_U(3) = X^{\langle 1 \rangle}_3+S_1 \\
    &A^{(1)}_U(4) = X^{\langle 1 \rangle}_1+X^{\langle 1 \rangle}_4+S_1,  &&A^{(2)}_U(4) = X^{\langle 1 \rangle}_4+S_1 \\
    &A^{(1)}_U(5) = X^{\langle 1 \rangle}_2+X^{\langle 1 \rangle}_3+S_1,  &&A^{(2)}_U(5) = X^{\langle 1 \rangle}_1+X^{\langle 1 \rangle}_2+X^{\langle 1 \rangle}_3+S_1 \\
    &A^{(1)}_U(6) = X^{\langle 1 \rangle}_2+X^{\langle 1 \rangle}_4+S_1,  &&A^{(2)}_U(6) = X^{\langle 1 \rangle}_1+X^{\langle 1 \rangle}_2+X^{\langle 1 \rangle}_4+S_1 \\
    &A^{(1)}_U(7) = X^{\langle 1 \rangle}_3+X^{\langle 1 \rangle}_4+S_1,  &&A^{(2)}_U(7) = X^{\langle 1 \rangle}_1+X^{\langle 1 \rangle}_3+X^{\langle 1 \rangle}_4+S_1 \\
    &A^{(1)}_U(8) = X^{\langle 1 \rangle}_1+X^{\langle 1 \rangle}_2+X^{\langle 1 \rangle}_3+X^{\langle 1 \rangle}_4+S_1,  &&A^{(2)}_U(8) = X^{\langle 1 \rangle}_2+X^{\langle 1 \rangle}_3+X^{\langle 1 \rangle}_4+S_1
\end{align}
Following the notation in Section~\ref{Problem Formulation}, the superscript of $A$ denotes the database index of $P_1$ while the index on the right-hand side of $A$ denotes the potential query choice that can be chosen by $P_2$. 

In order to retrieve $X^{\langle 1 \rangle}_2$, $P_2$ selects a random query choice for the first database of $P_1$ and its coupled query for the second database of $P_1$. For instance, $P_2$ chooses $1$ for the first database of $P_1$ and $2$ for the second database of $P_1$. After receiving the query symbol $1$, the first database of $P_1$ responds with the answer $A^{(1)}_U(1) = S_1$. Meanwhile, the second database of $P_1$ responds with the answer $A^{(2)}_U(2) = X^{\langle 1 \rangle}_2+S_1$ when the query symbol $2$ is received. Next, $P_2$ asks for the value of $X^{\langle 1 \rangle}_4$ from $P_1$ in the same way. 

Since there are 8 possible queries to each database of $P_1$, the communication cost from $P_2$ to $P_1$ is $3+3=6$ bits; and since each database of $P_1$ sends back a single bit of answer, the communication cost from $P_1$ to $P_2$ is $1+1=2$ bits. Thus, the total communication cost for learning $X^{\langle 1 \rangle}_2$ and $X^{\langle 1 \rangle}_4$ is $2\cdot (6+2)=16$ bits through this MM-SPIR approach. After learning the values of $X^{\langle 1 \rangle}_2$ and $X^{\langle 1 \rangle}_4$, $P_2$ knows that $P_1$ has element $2$ but does not have element $4$. Combining its own elements, $P_2$ is able to calculate the union, which is $\{1,2,3\}$. Thus, the PSU reliability constraint is satisfied. Regarding the two privacy constraints, due to the user privacy constraint in the SPIR problem, each individual database in $P_1$ can only learn that $P_2$ has two elements without learning any knowledge about what these two specific elements are. Due to the database privacy constraint in the SPIR problem, $P_2$ can only learn that $P_1$ possesses element $2$ and does not possess element $4$ without any additional knowledge about whether $P_1$ has elements $1, 3$. In particular, whether $P_1$ has element $4$ or not can be deduced by $P_2$ from the ultimate union result and its own elements. Thus, both of $P_1$ and $P_2$ privacy constraints are guaranteed. Thus, this is a valid two-party PSU scheme.
\end{example}

\begin{example} \label{two-party PSU2}
\textbf{Two-party PSU; one-database client; one-database leader:}
Compared to Example~\ref{two-party PSU1}, the only modification in the setting is that party $P_1$ now has a single database. Party $P_1$ also holds four server-side common randomness symbols $S_1$, $S_2$, $S_3$ and $S_4$ that are all uniformly selected from the finite field $\mathbb{F}_2$. In order to have a feasible single-database SPIR approach as illustrated in \cite{SPIR_atPIR}, the party $P_2$ obtains one random server-side common randomness symbol ahead of time. As a consequence, $P_1$ generates the following answer table for the only database, 
\begin{align}
    &A^{(1)}_U(1) = \{X^{\langle 1 \rangle}_1+S_1,X^{\langle 1 \rangle}_2+S_2,X^{\langle 1 \rangle}_3+S_3,X^{\langle 1 \rangle}_4+S_4\} \\
    &A^{(1)}_U(2) = \{X^{\langle 1 \rangle}_1+S_2,X^{\langle 1 \rangle}_2+S_3,X^{\langle 1 \rangle}_3+S_4,X^{\langle 1 \rangle}_4+S_1\} \\
    &A^{(1)}_U(3) = \{X^{\langle 1 \rangle}_1+S_3,X^{\langle 1 \rangle}_2+S_4,X^{\langle 1 \rangle}_3+S_1,X^{\langle 1 \rangle}_4+S_2\} \\
    &A^{(1)}_U(4) = \{X^{\langle 1 \rangle}_1+S_4,X^{\langle 1 \rangle}_2+S_1,X^{\langle 1 \rangle}_3+S_2,X^{\langle 1 \rangle}_4+S_3\}
\end{align}

Subsequently, $P_2$ selects a query choice that matches its pre-fetched server-side common randomness symbol. For instance, if its pre-fetched symbol is $S_1$, in order to retrieve $X^{\langle 1 \rangle}_2$, $P_2$ chooses $4$ and then sends this query symbol to $P_1$. The database belonging to $P_1$ responds with the answer $A^{(1)}_U(4) = \{X^{\langle 1 \rangle}_1+S_4,X^{\langle 1 \rangle}_2+S_1,X^{\langle 1 \rangle}_3+S_2,X^{\langle 1 \rangle}_4+S_3\}$. Likewise, $P_2$ also asks for the value of $X^{\langle 1 \rangle}_4$ from $P_1$ in the same way. Since there are 4 possible queries to the database of $P_1$, the communication cost from $P_2$ to $P_1$ is $2$ bits; and since $P_1$ sends back an answer with $4$ components, the communication cost from $P_1$ to $P_2$ is $4$ bits. Thus, the total communication cost for learning $X^{\langle 1 \rangle}_2$ and $X^{\langle 1 \rangle}_4$ is $2\cdot (2+4)=12$ bits through this MM-SPIR approach, without considering the communication cost generated by the pre-fetched server-side common randomness symbol. Verification that this achievable scheme satisfies the PSU reliability, $P_1$ privacy and $P_2$ privacy constraints follows similarly as in Example~\ref{two-party PSU1}.
\end{example}

\begin{example} \label{MP-PSU1}
\textbf{Five-party PSU; one-database per client; one-database leader:}
As a generalization of Examples~\ref{two-party PSU1} and \ref{two-party PSU2}, in this example, we consider a multi-party setting, again with the global alphabet $\mathcal{A} = \{1,2,3,4\}$. Here, the first party $P_1$ contains element $1$, i.e., $\mathcal{P}_1 = \{1\}$. The second party $P_2$ contains element 1 and element $3$, i.e., $\mathcal{P}_2 = \{1,3\}$. The third party $P_3$ contains element $1$ and element $4$, i.e., $\mathcal{P}_3 = \{1,4\}$. The fourth party $P_4$ contains element $1$, element $3$ and element $4$, i.e., $\mathcal{P}_4 = \{1,3,4\}$. The fifth party $P_5$ contains nothing, i.e., $\mathcal{P}_5 = \emptyset$. As before, we assume that the total number of elements in each party is public knowledge. The parties construct the corresponding incidence vectors $X^{\langle [5] \rangle}$ as follows,
\begin{align}
    &\mbox{Party} ~ P_1: \quad \mathcal{P}_1 = \{1\} \quad && \Rightarrow \quad X^{\langle 1 \rangle} = [X^{\langle 1 \rangle}_1 \:\: X^{\langle 1 \rangle}_2 \:\: X^{\langle 1 \rangle}_3 \:\: X^{\langle 1 \rangle}_4]^T = [1 \:\: 0 \:\: 0 \:\: 0]^T \\
    &\mbox{Party} ~ P_2: \quad \mathcal{P}_2 = \{1,3\} \quad && \Rightarrow \quad X^{\langle 2 \rangle} = [X^{\langle 2 \rangle}_1 \:\: X^{\langle 2 \rangle}_2 \:\: X^{\langle 2 \rangle}_3 \:\: X^{\langle 2 \rangle}_4]^T = [1 \:\: 0 \:\: 1 \:\: 0]^T \\
    &\mbox{Party} ~ P_3: \quad \mathcal{P}_3 = \{1,4\} \quad && \Rightarrow \quad X^{\langle 3 \rangle} = [X^{\langle 3 \rangle}_1 \:\: X^{\langle 3 \rangle}_2 \:\: X^{\langle 3 \rangle}_3 \:\: X^{\langle 3 \rangle}_4]^T = [1 \:\: 0 \:\: 0 \:\: 1]^T \\
    &\mbox{Party} ~ P_4: \quad \mathcal{P}_4 = \{1,3,4\} \quad && \Rightarrow \quad X^{\langle 4 \rangle} = [X^{\langle 4 \rangle}_1 \:\: X^{\langle 4 \rangle}_2 \:\: X^{\langle 4 \rangle}_3 \:\: X^{\langle 4 \rangle}_4]^T = [1 \:\: 0 \:\: 1 \:\: 1]^T \\
    &\mbox{Party} ~ P_5: \quad \mathcal{P}_5 = \emptyset \quad && \Rightarrow \quad X^{\langle 5 \rangle} = [X^{\langle 5 \rangle}_1 \:\: X^{\langle 5 \rangle}_2 \:\: X^{\langle 5 \rangle}_3 \:\: X^{\langle 5 \rangle}_4]^T = [0 \:\: 0 \:\: 0 \:\: 0]^T 
\end{align}

Using the MP-PSI achievable scheme in \cite{MP-PSI_journal} as a reference, we select party $P_5$ as the leader party, and the remaining parties as client parties, as party $P_5$ is globally known as an empty party. Thus, there is no need for $P_5$ to send any queries to the remaining parties and the server-side common randomness employed in the previous two examples is not necessary any more. In this example, all parties have a single database. Besides their own incidence vectors, each client party holds the same set of common randomness symbols\footnote{The common randomness symbol $u^{\langle i \rangle}_\alpha$ in this MP-PSU implementation functions exactly in the same way as the common randomness symbols $t_{i,j}$ functioned in the MP-PSI paper \cite{MP-PSI_journal}. The same is true for subsequent common randomness symbols $w^{\langle i \rangle}_\alpha$ in the write-back implementation.} $\{u^{\langle [4] \rangle}_\alpha \!\!: \alpha \in [4]\}$ from the finite field $\mathbb{F}_5$ as well as the same global common randomness symbol $c$ that is uniformly distributed over $\{1,2,3,4\}$. Moreover, $\{u^{\langle [4] \rangle}_\alpha \!\!: \alpha \in [4]\}$ are such that the sum $\sum_{i \in [4]} u^{\langle i \rangle}_\alpha$ is always equal to $0$ for all $\alpha \in [4]$. The answers from the client parties are,
\begin{align}
    &A_U^{\langle 1 \rangle} = \{c(X^{\langle 1 \rangle}_1+u^{\langle 1 \rangle}_1), c(X^{\langle 1 \rangle}_2+u^{\langle 1 \rangle}_2), c(X^{\langle 1 \rangle}_3+u^{\langle 1 \rangle}_3), c(X^{\langle 1 \rangle}_4+u^{\langle 1 \rangle}_4)\} \label{MP-PSU1 answer1} \\
    &A_U^{\langle 2 \rangle} = \{c(X^{\langle 2 \rangle}_1+u^{\langle 2 \rangle}_1), c(X^{\langle 2 \rangle}_2+u^{\langle 2 \rangle}_2), c(X^{\langle 2 \rangle}_3+u^{\langle 2 \rangle}_3), c(X^{\langle 2 \rangle}_4+u^{\langle 2 \rangle}_4)\} \label{MP-PSU1 answer2} \\
    &A_U^{\langle 3 \rangle} = \{c(X^{\langle 3 \rangle}_1+u^{\langle 3 \rangle}_1), c(X^{\langle 3 \rangle}_2+u^{\langle 3 \rangle}_2), c(X^{\langle 3 \rangle}_3+u^{\langle 3 \rangle}_3), c(X^{\langle 3 \rangle}_4+u^{\langle 3 \rangle}_4)\} \label{MP-PSU1 answer3} \\
    &A_U^{\langle 4 \rangle} = \{c(X^{\langle 4 \rangle}_1+u^{\langle 4 \rangle}_1), c(X^{\langle 4 \rangle}_2+u^{\langle 4 \rangle}_2), c(X^{\langle 4 \rangle}_3+u^{\langle 4 \rangle}_3), c(X^{\langle 4 \rangle}_4+u^{\langle 4 \rangle}_4)\} \label{MP-PSU1 answer4}
\end{align}

Regarding reliability: The leader party $P_5$ calculates the following expressions,
\begin{align}
    \mbox{Element} ~ 1: \quad &c(X^{\langle 1 \rangle}_1+u^{\langle 1 \rangle}_1) + c(X^{\langle 2 \rangle}_1+u^{\langle 2 \rangle}_1) +  c(X^{\langle 3 \rangle}_1+u^{\langle 3 \rangle}_1) + c(X^{\langle 4 \rangle}_1+u^{\langle 4 \rangle}_1) \notag \\
    &= c(\sum_{i \in [4]} X^{\langle i \rangle}_1 + \sum_{i \in [4]} u^{\langle i \rangle}_1) =  c(\sum_{i \in [4]} X^{\langle i \rangle}_1) = c \cdot 4 \neq 0 \label{MP-PSU1 analysis1} \\
    \mbox{Element} ~ 2: \quad &c(X^{\langle 1 \rangle}_2+u^{\langle 1 \rangle}_2) + c(X^{\langle 2 \rangle}_2+u^{\langle 2 \rangle}_2) +  c(X^{\langle 3 \rangle}_2+u^{\langle 3 \rangle}_2) + c(X^{\langle 4 \rangle}_2+u^{\langle 4 \rangle}_2) \notag \\
    &= c(\sum_{i \in [4]} X^{\langle i \rangle}_2 + \sum_{i \in [4]} u^{\langle i \rangle}_2) =  c(\sum_{i \in [4]} X^{\langle i \rangle}_2) = c \cdot 0 = 0 \label{MP-PSU1 analysis2} \\
    \mbox{Element} ~ 3: \quad &c(X^{\langle 1 \rangle}_3+u^{\langle 1 \rangle}_3) + c(X^{\langle 2 \rangle}_3+u^{\langle 2 \rangle}_3) +  c(X^{\langle 3 \rangle}_3+u^{\langle 3 \rangle}_3) + c(X^{\langle 4 \rangle}_3+u^{\langle 4 \rangle}_3) \notag \\
    &= c(\sum_{i \in [4]} X^{\langle i \rangle}_3 + \sum_{i \in [4]} u^{\langle i \rangle}_3) =  c(\sum_{i \in [4]} X^{\langle i \rangle}_3) = c \cdot 2 \neq 0 \label{MP-PSU1 analysis3} \\
    \mbox{Element} ~ 4: \quad &c(X^{\langle 1 \rangle}_4+u^{\langle 1 \rangle}_4) + c(X^{\langle 2 \rangle}_4+u^{\langle 2 \rangle}_4) +  c(X^{\langle 3 \rangle}_4+u^{\langle 3 \rangle}_4) + c(X^{\langle 4 \rangle}_4+u^{\langle 4 \rangle}_4) \notag \\
    &= c(\sum_{i \in [4]} X^{\langle i \rangle}_4 + \sum_{i \in [4]} u^{\langle i \rangle}_4) =  c(\sum_{i \in [4]} X^{\langle i \rangle}_4) = c \cdot 2 \neq 0 \label{MP-PSU1 analysis4}
\end{align}
Thus, $P_5$ concludes that the union $\mathcal{P}_1 \cup \mathcal{P}_2 \cup \mathcal{P}_3 \cup \mathcal{P}_4$ is $\{1,3,4\}$ because the first, third and fourth expressions are not equal to $0$.

Regarding privacy: The leader party's privacy constraint is trivially satisfied since the leader party is empty and sent no queries to the clients. The clients' privacies are protected by the common randomness symbols $\{u^{\langle [4] \rangle}_\alpha \!\!: \alpha \in [4]\}$ and $c$. First, the values of the individual components of the incidence vector $\{X^{\langle [4] \rangle}_\alpha \!\!: \alpha \in [4]\}$ are kept private from $P_5$ by the added randomness symbols $\{u^{\langle [4] \rangle}_\alpha \!\!: \alpha \in [4]\}$. These \emph{coupled} (i.e., correlated) random variables disappear when the components coming from clients are added up as $\sum_{i \in [4]} u^{\langle i \rangle}_\alpha$ is always $0$ for all $\alpha \in [4]$. Finally, the global common randomness symbol $c$ protects the actual value of the sum $\sum_i X^{\langle i \rangle}_\alpha$ for all $\alpha$. That is, leader $P_5$ can only learn whether these sums are zero or not and nothing beyond that. Thus, this is a valid scheme satisfying reliability and privacy.
\end{example}

\begin{example} \label{MP-PSU2}
\textbf{Five-party PSU; one-database per client; two-database leader:}
With respect to the MP-PSU configuration in Example~\ref{MP-PSU1}, we only change the number of databases in the leader party $P_5$, which now contains two replicated and non-colluding databases. As these two databases do not communicate with each other directly, a straightforward approach could be to have each client send its answer shown in \eqref{MP-PSU1 answer1}-\eqref{MP-PSU1 answer4} to both databases in the leader party. This way, each database could individually learn the union while the privacy constraints are still satisfied. Here, we put forth an alternative approach, where two random client parties are utilized as intermediaries to route the information between the two non-colluding databases in the leader party such that there is no need for a client to send the replicated answer to both databases in $P_5$. To that end, each client party also holds another duplicate set of common randomness symbols $\{u_\alpha \!\!: \alpha \in [4]\}$ that are all uniformly selected from $\mathbb{F}_5$ on the basis of the existing common randomness symbols $\{u^{\langle [4] \rangle}_\alpha \!\!: \alpha \in [4]\}$. Since $P_5$ does not have any element, there is no need for the other parties to download any information from $P_5$ in the beginning, i.e., $D_{U,1}^{\langle \mathcal{C}_1 \rangle,(1)}$ and $D_{U,1}^{\langle \mathcal{C}_2 \rangle,(2)}$ are both null. At this point, let the client parties $P_1$ and $P_2$ form the first group. They send their respective answers $A_{U,1}^{\langle 1 \rangle,(1)}$ and $A_{U,1}^{\langle 2 \rangle,(1)}$ as shown in \eqref{MP-PSU1 answer1}-\eqref{MP-PSU1 answer2} to the first database of $P_5$ since they are associated with database $1$. This database produces a response $D_{U,2}^{\langle 2 \rangle,(1)}$ to be downloaded by client $2$ through element-wisely adding its received answers and appending leader party common randomness symbols that are all uniformly selected from the finite field $\mathbb{F}_5$, 
\begin{align}
    D_{U,2}^{\langle 2 \rangle,(1)} = \{&c(X^{\langle 1 \rangle}_1+X^{\langle 2 \rangle}_1+u^{\langle 1 \rangle}_1+u^{\langle 2 \rangle}_1)+S_1,c(X^{\langle 1 \rangle}_2+X^{\langle 2 \rangle}_2+u^{\langle 1 \rangle}_2+u^{\langle 2 \rangle}_2)+S_2, \notag \\
    &c(X^{\langle 1 \rangle}_3+X^{\langle 2 \rangle}_3+u^{\langle 1 \rangle}_3+u^{\langle 2 \rangle}_3)+S_3,c(X^{\langle 1 \rangle}_4+X^{\langle 2 \rangle}_4+u^{\langle 1 \rangle}_4+u^{\langle 2 \rangle}_4)+S_4\}
\end{align}
This database then sends this response back to $P_2$. After adding extra common randomness to the received response, $P_2$ forwards the following answer to both databases in $P_5$, 
\begin{align}
    A_{U,2}^{\langle 2 \rangle,([2])} = \{&c(X^{\langle 1 \rangle}_1+X^{\langle 2 \rangle}_1+u^{\langle 1 \rangle}_1+u^{\langle 2 \rangle}_1)+u_1+S_1,c(X^{\langle 1 \rangle}_2+X^{\langle 2 \rangle}_2+u^{\langle 1 \rangle}_2+u^{\langle 2 \rangle}_2)+u_2+S_2, \notag \\
    &c(X^{\langle 1 \rangle}_3+X^{\langle 2 \rangle}_3+u^{\langle 1 \rangle}_3+u^{\langle 2 \rangle}_3)+u_3+S_3,c(X^{\langle 1 \rangle}_4+X^{\langle 2 \rangle}_4+u^{\langle 1 \rangle}_4+u^{\langle 2 \rangle}_4)+u_4+S_4\}
\end{align}
Meanwhile, the client parties $P_3$ and $P_4$, which form the second group, send their respective answers $A_{U,1}^{\langle 3 \rangle,(2)}$ and $A_{U,1}^{\langle 4 \rangle,(2)}$ as shown in \eqref{MP-PSU1 answer3}-\eqref{MP-PSU1 answer4} to the second database of $P_5$. Similarly, this database produces a response $D_{U,2}^{\langle 3 \rangle,(2)}$ to be downloaded by client $3$ as follows, and sends it back to $P_3$,
\begin{align}
    D_{U,2}^{\langle 3 \rangle,(2)} = \{&c(X^{\langle 3 \rangle}_1+X^{\langle 4 \rangle}_1+u^{\langle 3 \rangle}_1+u^{\langle 4 \rangle}_1)-S_1,c(X^{\langle 3 \rangle}_2+X^{\langle 4 \rangle}_2+u^{\langle 3 \rangle}_2+u^{\langle 4 \rangle}_2)-S_2, \notag \\
    &c(X^{\langle 3 \rangle}_3+X^{\langle 4 \rangle}_3+u^{\langle 3 \rangle}_3+u^{\langle 4 \rangle}_3)-S_3,c(X^{\langle 3 \rangle}_4+X^{\langle 4 \rangle}_4+u^{\langle 3 \rangle}_4+u^{\langle 4 \rangle}_4)-S_4\}
\end{align}
Then, $P_3$ forwards the following further processed answer to both databases in $P_5$,
\begin{align}
    A_{U,2}^{\langle 3 \rangle,([2])} = \{&c(X^{\langle 3 \rangle}_1+X^{\langle 4 \rangle}_1+u^{\langle 3 \rangle}_1+u^{\langle 4 \rangle}_1)-u_1-S_1,c(X^{\langle 3 \rangle}_2+X^{\langle 4 \rangle}_2+u^{\langle 3 \rangle}_2+u^{\langle 4 \rangle}_2)-u_2-S_2, \notag \\
    &c(X^{\langle 3 \rangle}_3+X^{\langle 4 \rangle}_3+u^{\langle 3 \rangle}_3+u^{\langle 4 \rangle}_3)-u_3-S_3,c(X^{\langle 3 \rangle}_4+X^{\langle 4 \rangle}_4+u^{\langle 3 \rangle}_4+u^{\langle 4 \rangle}_4)-u_4-S_4\}
\end{align}
After collecting the answers in the second communication step, each individual database $j$ in $P_5$ finds the desired submodel union by element-wisely adding $A_{U,2}^{\langle 2 \rangle,(j)}$ and $A_{U,2}^{\langle 3 \rangle,(j)}$,
\begin{align}
    A_{U,2}^{\langle 2 \rangle,(j)} + A_{U,2}^{\langle 3 \rangle,(j)} = \biggl\{c(\sum_{i \in [4]} X^{\langle i \rangle}_1), c(\sum_{i \in [4]} X^{\langle i \rangle}_2), c(\sum_{i \in [4]} X^{\langle i \rangle}_3), c(\sum_{i \in [4]} X^{\langle i \rangle}_4)\biggr\}, \quad \forall j
\end{align}

Regarding reliability: The MP-PSU reliability in the leader party $P_5$ is inherited from the MP-PSU reliability analysis in Example~\ref{MP-PSU1}. Specifically, each individual database in $P_5$ can make the same analysis as shown in \eqref{MP-PSU1 analysis1}-\eqref{MP-PSU1 analysis4} for each element in the alphabet to derive the union $\mathcal{P}_1 \cup \mathcal{P}_2 \cup \mathcal{P}_3 \cup \mathcal{P}_4$. Also, $P_5$ can send this union result to any client party if needed. 

Regarding privacy: The privacy analysis of the client parties $P_1$ and $P_4$ is trivial, since neither of them has received any information from the remaining parties. Regarding the client party $P_2$, due to the appended leader party common randomness $\{S_{\alpha}, \alpha \in [4]\}$, this party cannot learn anything about the incidence vector symbols in the remaining parties from its only received information $D_{U,2}^{\langle 2 \rangle,(1)}$. This analysis also applies to the client party $P_3$. Regarding the leader party $P_5$, it is obvious that the received information $\{A_{U,1}^{\langle 1 \rangle,(1)},A_{U,1}^{\langle 2 \rangle,(1)},A_{U,2}^{\langle 2 \rangle,(1)},A_{U,2}^{\langle 3 \rangle,(1)}\}$ in the first database and the received information $\{A_{U,1}^{\langle 3 \rangle,(2)},A_{U,1}^{\langle 4 \rangle,(2)},A_{U,2}^{\langle 2 \rangle,(2)},A_{U,2}^{\langle 3 \rangle,(2)}\}$ in the second database, individually, contain less information about the incidence vector symbols in the client parties than the answer set $A_U^{\langle [4] \rangle}$ received by $P_5$ in Example~\ref{MP-PSU1}. Therefore, the leader party $P_5$ can only learn the union and nothing beyond that. Thus, this is a valid MP-PSU scheme. 

Next, we consider situations that are commonly encountered in practical implementations.

First, one or more client parties may drop-out during the MP-PSU process. For instance, $P_1$ may lose connection to $P_5$, in which case, the first database of $P_5$ will only receive the answer from $P_2$. The download produced in the original method now becomes,
\begin{align}
    D_{U,2}^{\prime\langle 2 \rangle,(1)} = \{&c(X^{\langle 2 \rangle}_1+u^{\langle 2 \rangle}_1)+S_1,c(X^{\langle 2 \rangle}_2+u^{\langle 2 \rangle}_2)+S_2, \notag \\
    &c(X^{\langle 2 \rangle}_3+u^{\langle 2 \rangle}_3)+S_3,c(X^{\langle 2 \rangle}_4+u^{\langle 2 \rangle}_4)+S_4\}
\end{align}
It is easy to observe that the common randomness symbols $u^{\langle [4] \rangle}_\alpha$ in these two downloads cannot be cancelled completely as before. However, note that $P_2$ possesses the missing common randomness symbols incurred by $P_1$ drop-out. Hence, $P_2$ can add the required common randomness itself as long as it learns from database $1$ that $P_1$ has dropped-out. Thus, the adjusted answer in the second step $A_{U,2}^{\prime\langle 2 \rangle,([2])}$ is as follows and will be sent back to both databases in $P_5$,
\begin{align}
    A_{U,2}^{\prime\langle 2 \rangle,([2])} = \{&c(X^{\langle 2 \rangle}_1+u^{\langle 1 \rangle}_1+u^{\langle 2 \rangle}_1)+u_1+S_1,c(X^{\langle 2 \rangle}_2+u^{\langle 1 \rangle}_2+u^{\langle 2 \rangle}_2)+u_2+S_2, \notag \\
    &c(X^{\langle 2 \rangle}_3+u^{\langle 1 \rangle}_3+u^{\langle 2 \rangle}_3)+u_3+S_3,c(X^{\langle 2 \rangle}_4+u^{\langle 1 \rangle}_4+u^{\langle 2 \rangle}_4)+u_4+S_4\}
\end{align}
Further, if $P_2$ loses its connection to $P_5$, the remaining active client party $P_1$ in the first client party group functions as a router. Since no one in the second client group drops-out, the download $D_{U,2}^{\langle 3 \rangle,(2)}$ remains the same,
\begin{align}
    D_{U,2}^{\langle 3 \rangle,(2)} = \{&c(X^{\langle 3 \rangle}_1+X^{\langle 4 \rangle}_1+u^{\langle 3 \rangle}_1+u^{\langle 4 \rangle}_1)-S_1,c(X^{\langle 3 \rangle}_2+X^{\langle 4 \rangle}_2+u^{\langle 3 \rangle}_2+u^{\langle 4 \rangle}_2)-S_2, \notag \\
    &c(X^{\langle 3 \rangle}_3+X^{\langle 4 \rangle}_3+u^{\langle 3 \rangle}_3+u^{\langle 4 \rangle}_3)-S_3,c(X^{\langle 3 \rangle}_4+X^{\langle 4 \rangle}_4+u^{\langle 3 \rangle}_4+u^{\langle 4 \rangle}_4)-S_4\}
\end{align}
The result forwarded by $P_3$ and received by the databases in $P_5$ remains the same,
\begin{align}
    A_{U,2}^{\langle 3 \rangle,([2])} = \{&c(X^{\langle 3 \rangle}_1+X^{\langle 4 \rangle}_1+u^{\langle 3 \rangle}_1+u^{\langle 4 \rangle}_1)-u_1-S_1,c(X^{\langle 3 \rangle}_2+X^{\langle 4 \rangle}_2+u^{\langle 3 \rangle}_2+u^{\langle 4 \rangle}_2)-u_2-S_2, \notag \\
    &c(X^{\langle 3 \rangle}_3+X^{\langle 4 \rangle}_3+u^{\langle 3 \rangle}_3+u^{\langle 4 \rangle}_3)-u_3-S_3,c(X^{\langle 3 \rangle}_4+X^{\langle 4 \rangle}_4+u^{\langle 3 \rangle}_4+u^{\langle 4 \rangle}_4)-u_4-S_4\}
\end{align}
We can now verify that both databases in $P_5$ can determine the union $\mathcal{P}_2 \cup \mathcal{P}_3 \cup \mathcal{P}_4$ without the participation of $P_1$. 

Second, the answer $A_{U,1}^{\langle 1 \rangle,(1)}$ generated by $P_1$ may arrive at database $1$ in $P_5$ so late that database $1$ may believe that $P_1$ has dropped-out. In such a case, the privacy in our MP-PSU is still preserved. If we look at the received information $\{A_{U,1}^{\langle 1 \rangle,(1)},A_{U,1}^{\langle 2 \rangle,(1)},A_{U,2}^{\prime\langle 2 \rangle,(1)},A_{U,2}^{\langle 3 \rangle,(1)}\}$ in database $1$ of $P_5$, no information about the incidence vector $X^{\langle 1 \rangle}$ is leaked due to the existence of extra common randomness symbols $\{u_\alpha \!\!: \alpha \in [4]\}$. Moreover, this late answer $A_{U,1}^{\langle 1 \rangle,(1)}$ will never be transmitted to any other client parties by $P_5$ in order to avoid the further leakage of $X^{\langle 1 \rangle}$. The usage of extra common randomness $u_\alpha$ here is similar to the double-masking idea in \cite{SecAgg} so as to resolve this late arrival problem, but in a very simple manner.

Third, one of the two databases in $P_5$ may also drop-out during the implementation. For instance, if database $2$ drops-out, the same answers $\{A_{U,1}^{\langle 1 \rangle,(1)},A_{U,1}^{\langle 2 \rangle,(1)},A_{U,2}^{\langle 2 \rangle,(1)}\}$ can still be received by database $1$ in $P_5$ from $P_1$ and $P_2$, but  $\{A_{U,2}^{\langle 3 \rangle,(1)}\}$ cannot be received from the other client party group as usual. The corresponding remedy is that the surviving database asks for the values of $\{c(u^{\langle 3 \rangle}_\alpha + u^{\langle 4 \rangle}_\alpha) \!\!: \alpha \in [4]\}$ from $P_2$ through one more communication round. In this way, it is easy to see that the first database in $P_5$ can derive the union $\mathcal{P}_1 \cup \mathcal{P}_2$ associated with the first client party group.
\end{example}

\begin{example} \label{FSL}
\textbf{Five-party PSU; one-database per client; two-database leader; together with FSL-write:}
Consider a distributed FSL problem involving a server consisting of two replicated and non-colluding databases and four selected clients in this round of FSL process. Each individual database stores $4$ independent submodels each containing 2 i.i.d.~symbols uniformly selected from a sufficiently large finite field $\mathbb{F}_q$, $q \geq 5$, i.e., $M_1 = [M_{1,1},M_{1,2}], M_2 = [M_{2,1},M_{2,2}], M_3 = [M_{3,1},M_{3,2}], M_4 = [M_{4,1},M_{4,2}]$ and some required server-side common randomness symbols. According to the clients' respective local training data, client $1$ can be used to update the submodel $1$, client $2$ can be used to update the submodels $1,3$, client $3$ can be used to update the submodels $1,4$ and client $4$ can be used to update the submodels $1,3,4$, i.e., $\Gamma^{\langle 1 \rangle} = \{1\}, \Gamma^{\langle 2 \rangle} = \{1,3\}, \Gamma^{\langle 3 \rangle} = \{1,4\}, \Gamma^{\langle 4 \rangle} = \{1,3,4\}$. Both databases in the server can communicate with each client through a secure and authenticated channel. We further assume that the channels connected to database $1$ have higher bandwidth than the ones connected to database $2$ for clients $1,2$ and it is the opposite for clients $3,4$. Thus, the FSL-PSU phase is executed exactly as in the MP-PSU in Example~\ref{MP-PSU2}, and each database in the server learns the desired submodel union $\Gamma = \{1,3,4\}$ when this phase is complete. 

Due to the similarities between the formulations of FSL-PSU phase and the FSL-write phase, we use the idea in Example~\ref{MP-PSU2} one more time to execute the  FSL-write phase. Database $1$ sends the submodels $1,3,4$ to client $1$ and client $2$, while database $2$ sends the submodels $1,3,4$ to client $3$ and client $4$, i.e., the downloads $D^{\langle 1,2 \rangle,(1)}_{W,1}$ and $D^{\langle 3,4 \rangle,(2)}_{W,1}$ are both $\{M_1,M_3,M_4\}$. After receiving the desired submodels from the server, client $1$ generates the increment $\big\{\Delta^{\langle 1 \rangle}_{1,1}, \Delta^{\langle 1 \rangle}_{1,2}\big\}$ for submodel $1$, client $2$ generates the increment $\big\{\Delta^{\langle 2 \rangle}_{1,1}, \Delta^{\langle 2 \rangle}_{1,2}, \Delta^{\langle 2 \rangle}_{3,1}, \Delta^{\langle 2 \rangle}_{3,2}\big\}$ for submodels $1,3$,  client $3$ generates the increment $\big\{\Delta^{\langle 3 \rangle}_{1,1}, \Delta^{\langle 3 \rangle}_{1,2}, \Delta^{\langle 3 \rangle}_{4,1}, \Delta^{\langle 3 \rangle}_{4,2}\big\}$ for submodels $1,4$,  client $4$ generates the increment $\big\{\Delta^{\langle 4 \rangle}_{1,1}, \Delta^{\langle 4 \rangle}_{1,2}, \Delta^{\langle 4 \rangle}_{3,1}, \Delta^{\langle 4 \rangle}_{3,2}, \Delta^{\langle 4 \rangle}_{4,1}, \Delta^{\langle 4 \rangle}_{4,2}\big\}$ for submodels $1,3,4$ after performing their respective local training. In addition, we assume that each client has already obtained two sets of common randomness symbols $\{w^{\langle \Gamma \rangle}_{k,l} \!\!: k \in [4], l \in [2]\}$ and $\{w_{k,l} \!\!: k \in [4], l \in [2]\}$ from the finite field $\mathbb{F}_q$ as in the previous examples. For all $k \in \Gamma$ and all $l \in [2]$, the sum $\sum_{i \in [4]} w^{\langle i \rangle}_{k,l}$ is always equal to $0$. Thus, the answers sent to database $1$ in the server from clients $1$ and $2$ are as follows,
\begin{align}
    &A^{\langle 1 \rangle,(1)}_{W,1} = \{\Delta^{\langle 1 \rangle}_{1,1} + w^{\langle 1 \rangle}_{1,1}, \Delta^{\langle 1 \rangle}_{1,2} + w^{\langle 1 \rangle}_{1,2}, w^{\langle 1 \rangle}_{3,1}, w^{\langle 1 \rangle}_{3,2}, w^{\langle 1 \rangle}_{4,1}, w^{\langle 1 \rangle}_{4,2}\} \\
    &A^{\langle 2 \rangle,(1)}_{W,1} = \{\Delta^{\langle 2 \rangle}_{1,1} + w^{\langle 2 \rangle}_{1,1}, \Delta^{\langle 2 \rangle}_{1,2} + w^{\langle 2 \rangle}_{1,2}, \Delta^{\langle 2 \rangle}_{3,1} + w^{\langle 2 \rangle}_{3,1}, \Delta^{\langle 2 \rangle}_{3,2} + w^{\langle 2 \rangle}_{3,2}, w^{\langle 2 \rangle}_{4,1}, w^{\langle 2 \rangle}_{4,2}\}
\end{align}
After collecting the answers from clients $1,2$, database $1$ performs the element-wise summation with the aid of its own server-side common randomness symbols and transmits the following response to client $2$ in its associated client group,
\begin{align}
    D^{\langle 2 \rangle,(1)}_{W,2} = \{&\Delta^{\langle 1 \rangle}_{1,1} + \Delta^{\langle 2 \rangle}_{1,1} + w^{\langle 1 \rangle}_{1,1} + w^{\langle 2 \rangle}_{1,1} + S_{1,1}, \Delta^{\langle 1 \rangle}_{1,2} + \Delta^{\langle 2 \rangle}_{1,2} + w^{\langle 1 \rangle}_{1,2} + w^{\langle 2 \rangle}_{1,2} + S_{1,2}, \nonumber\\
    &\Delta^{\langle 2 \rangle}_{3,1} + w^{\langle 1 \rangle}_{3,1} + w^{\langle 2 \rangle}_{3,1} + S_{3,1}, \Delta^{\langle 2 \rangle}_{3,2} + w^{\langle 1 \rangle}_{3,2} + w^{\langle 2 \rangle}_{3,2} + S_{3,2}, \nonumber\\
    &w^{\langle 1 \rangle}_{4,1} + w^{\langle 2 \rangle}_{4,1} + S_{4,1}, w^{\langle 1 \rangle}_{4,2} + w^{\langle 2 \rangle}_{4,2} + S_{4,2}\}
\end{align}
Afterwards, client $2$ processes the received response by adding extra common randomness and then forwards the following answer to both databases in the server,
\begin{align}
    A^{\langle 2 \rangle,([2])}_{W,2} = \{&\Delta^{\langle 1 \rangle}_{1,1} + \Delta^{\langle 2 \rangle}_{1,1} + w^{\langle 1 \rangle}_{1,1} + w^{\langle 2 \rangle}_{1,1} + w_{1,1} +S_{1,1}, \Delta^{\langle 1 \rangle}_{1,2} + \Delta^{\langle 2 \rangle}_{1,2} + w^{\langle 1 \rangle}_{1,2} + w^{\langle 2 \rangle}_{1,2} + w_{1,2} + S_{1,2}, \nonumber\\
    &\Delta^{\langle 2 \rangle}_{3,1} + w^{\langle 1 \rangle}_{3,1} + w^{\langle 2 \rangle}_{3,1} + w_{3,1} + S_{3,1}, \Delta^{\langle 2 \rangle}_{3,2} + w^{\langle 1 \rangle}_{3,2} + w^{\langle 2 \rangle}_{3,2} + w_{3,2} + S_{3,2}, \nonumber\\
    &w^{\langle 1 \rangle}_{4,1} + w^{\langle 2 \rangle}_{4,1} + w_{4,1} + S_{4,1}, w^{\langle 1 \rangle}_{4,2} + w^{\langle 2 \rangle}_{4,2} + w_{4,2} + S_{4,2}\}
\end{align}
At the same time, the answers sent to database $2$ in the server from clients $3$ and $4$ are,
\begin{align}
    &A^{\langle 3 \rangle,(2)}_{W,1} = \{\Delta^{\langle 3 \rangle}_{1,1} + w^{\langle 3 \rangle}_{1,1}, \Delta^{\langle 3 \rangle}_{1,2} + w^{\langle 3 \rangle}_{1,2}, w^{\langle 3 \rangle}_{3,1}, w^{\langle 3 \rangle}_{3,2}, \Delta^{\langle 3 \rangle}_{4,1} + w^{\langle 3 \rangle}_{4,1}, \Delta^{\langle 3 \rangle}_{4,2} + w^{\langle 3 \rangle}_{4,2}\} \\
    &A^{\langle 4 \rangle,(2)}_{W,1} = \{\Delta^{\langle 4 \rangle}_{1,1} + w^{\langle 4 \rangle}_{1,1}, \Delta^{\langle 4 \rangle}_{1,2} + w^{\langle 4 \rangle}_{1,2}, \Delta^{\langle 4 \rangle}_{3,1} + w^{\langle 4 \rangle}_{3,1}, \Delta^{\langle 4 \rangle}_{3,2} + w^{\langle 4 \rangle}_{3,2}, \Delta^{\langle 4 \rangle}_{4,1} + w^{\langle 4 \rangle}_{4,1}, \Delta^{\langle 4 \rangle}_{4,2} + w^{\langle 4 \rangle}_{4,2}\}
\end{align}
When the collection and computation is finished, database $2$ sends the following response to client $3$ who belongs to its associated client group,
\begin{align}
    D^{\langle 3 \rangle,(2)}_{W,2} = \{&\Delta^{\langle 3 \rangle}_{1,1} + \Delta^{\langle 4 \rangle}_{1,1} + w^{\langle 3 \rangle}_{1,1} + w^{\langle 4 \rangle}_{1,1} - S_{1,1}, \Delta^{\langle 3 \rangle}_{1,2} + \Delta^{\langle 4 \rangle}_{1,2} + w^{\langle 3 \rangle}_{1,2} + w^{\langle 4 \rangle}_{1,2} - S_{1,2}, \nonumber\\
    &\Delta^{\langle 4 \rangle}_{3,1} + w^{\langle 3 \rangle}_{3,1} + w^{\langle 4 \rangle}_{3,1} - S_{3,1}, \Delta^{\langle 4 \rangle}_{3,2} + w^{\langle 3 \rangle}_{3,2} + w^{\langle 4 \rangle}_{3,2} - S_{3,2}, \nonumber\\
    &\Delta^{\langle 3 \rangle}_{4,1} + \Delta^{\langle 4 \rangle}_{4,1} + w^{\langle 3 \rangle}_{4,1} + w^{\langle 4 \rangle}_{4,1} - S_{4,1}, \Delta^{\langle 3 \rangle}_{4,2} + \Delta^{\langle 4 \rangle}_{4,2} + w^{\langle 3 \rangle}_{4,2} + w^{\langle 4 \rangle}_{4,2} - S_{4,2}\}
\end{align}
Similarly, client $3$ processes the received response by adding extra common randomness again and then forwards the following answer to both databases in the server,
\begin{align}
    A^{\langle 3 \rangle,([2])}_{W,2} = \{&\Delta^{\langle 3 \rangle}_{1,1} + \Delta^{\langle 4 \rangle}_{1,1} + w^{\langle 3 \rangle}_{1,1} + w^{\langle 4 \rangle}_{1,1} - w_{1,1} - S_{1,1}, \Delta^{\langle 3 \rangle}_{1,2} + \Delta^{\langle 4 \rangle}_{1,2} + w^{\langle 3 \rangle}_{1,2} + w^{\langle 4 \rangle}_{1,2} - w_{1,2} - S_{1,2}, \nonumber\\
    &\Delta^{\langle 4 \rangle}_{3,1} + w^{\langle 3 \rangle}_{3,1} + w^{\langle 4 \rangle}_{3,1} - w_{3,1} - S_{3,1}, \Delta^{\langle 4 \rangle}_{3,2} + w^{\langle 3 \rangle}_{3,2} + w^{\langle 4 \rangle}_{3,2} - w_{3,2} - S_{3,2}, \nonumber\\
    &\Delta^{\langle 3 \rangle}_{4,1} + \Delta^{\langle 4 \rangle}_{4,1} + w^{\langle 3 \rangle}_{4,1} + w^{\langle 4 \rangle}_{4,1} - w_{4,1} - S_{4,1}, \Delta^{\langle 3 \rangle}_{4,2} + \Delta^{\langle 4 \rangle}_{4,2} + w^{\langle 3 \rangle}_{4,2} + w^{\langle 4 \rangle}_{4,2} - w_{4,2} - S_{4,2}\}
\end{align}
At this point, both databases can update the corresponding submodels after receiving the answers in the second step and removing all the involved common randomness symbols through element-wise summation,
\begin{align}
    M_1^{\prime} &= \{M_{1,1} + \Delta^{\langle 1 \rangle}_{1,1} + \Delta^{\langle 2 \rangle}_{1,1} + \Delta^{\langle 3 \rangle}_{1,1} + \Delta^{\langle 4 \rangle}_{1,1}, M_{1,2} + \Delta^{\langle 1 \rangle}_{1,2} + \Delta^{\langle 2 \rangle}_{1,2} + \Delta^{\langle 3 \rangle}_{1,2} + \Delta^{\langle 4 \rangle}_{1,2}\} \\
    M_3^{\prime} &= \{M_{3,1} + \Delta^{\langle 2 \rangle}_{3,1} + \Delta^{\langle 4 \rangle}_{3,1}, M_{3,2} + \Delta^{\langle 2 \rangle}_{3,2} + \Delta^{\langle 4 \rangle}_{3,2}\} \\
    M_4^{\prime} &= \{M_{4,1} + \Delta^{\langle 3 \rangle}_{4,1} + \Delta^{\langle 4 \rangle}_{4,1}, M_{4,2} + \Delta^{\langle 3 \rangle}_{4,2} + \Delta^{\langle 4 \rangle}_{4,2}\} 
\end{align}

In this example, we note that the scheme used in the FSL-write phase is a simplified version of the MP-PSU scheme used in Example~\ref{MP-PSU2} without considering the global common randomness symbol $c$. Therefore, regarding this FSL-write scheme, we can readily verify the FSL-write reliability constraint as well as the FSL-write privacy constraint for each individual database at the server side in reference to the leader party $P_5$ in Example~\ref{MP-PSU2}, and the FSL-write inter-client privacy constraint for clients $2,3$ in reference to the client parties $P_2, P_3$ in Example~\ref{MP-PSU2}. Likewise, the robustness against client drop-outs, client late arrivals and database drop-outs possessed by this FSL-write scheme is also inherited from the one in Example~\ref{MP-PSU2}. Finally, as required by the FSL process itself, the FSL-PSU phase and FSL-write phase introduced in this example can be executed repeatedly if a new set of clients are selected to perform another round of this FSL process. New sets of common randomness symbols and server-side common randomness symbols are needed to ensure privacy in each round.
\end{example}

\section{General FSL Achievable Scheme} \label{General FSL Achievable Scheme}
In this section, we describe our general achievable scheme for a distributed FSL model with any arbitrary initial parameters; see the model formulation in Section~\ref{FSL Formulation}. Our general achievable scheme has three phases: common randomness generation phase (FSL-CRG), private determination of the union of indices of submodels to be updated (FSL-PSU), and private writing of the updated submodels in the union back to the databases (FSL-write). In Section~\ref{Example Section}, we have given examples of FSL-PSU, and combined FSL-PSU and FSL-write. The FSL-PSU and FSL-write phases make use of pre-established common randomness at the client side. In Section~\ref{Example Section}, we have presumed that the common randomness needed for FSL-PSU and FSL-write have already been established. In this section, we first describe the establishment of the necessary common randomness across the clients. Our FSL-CRG scheme exploits the distributed nature of the server databases, and uses one-time pads and the RSPIR scheme introduced in \cite{RSPIR} to generate common randomness. Then, we describe FSL-PSU and FSL-write for the most general case.

\subsection{Common Randomness Generation (FSL-CRG) Phase} \label{Common Randomness Allocation Phase}
The two databases in the server aim to establish two types of common randomness across the clients: The first type is a global common randomness symbol $c$ that is uniformly selected from the set $\mathbb{F}_q \backslash \{0\}$. The second type is a set of general common randomness symbols $\{R_0,R_1,\dots,R_C\}$ with a flexible set length $C+1$, where each symbol is uniformly selected from $\mathbb{F}_q$ and the sum of the last $C$ symbols is equal to $0$, i.e., $\sum_{i \in [C]} R_i = 0$. As a result, $R_0$ can be used as $u_k$ or $w_{k,l}$ while $R_{[C]}$ can be used as $u^{\langle [C] \rangle}_k$ or $w^{\langle [C] \rangle}_{k,l}$ in the next two phases. The FSL-CRG phase is independent of the FSL-PSU and FSL-write phases in a practical implementation, and therefore, can be potentially executed during the off-peak hours. 

\begin{figure}[t]
\centering
\includegraphics[width=0.65\linewidth]{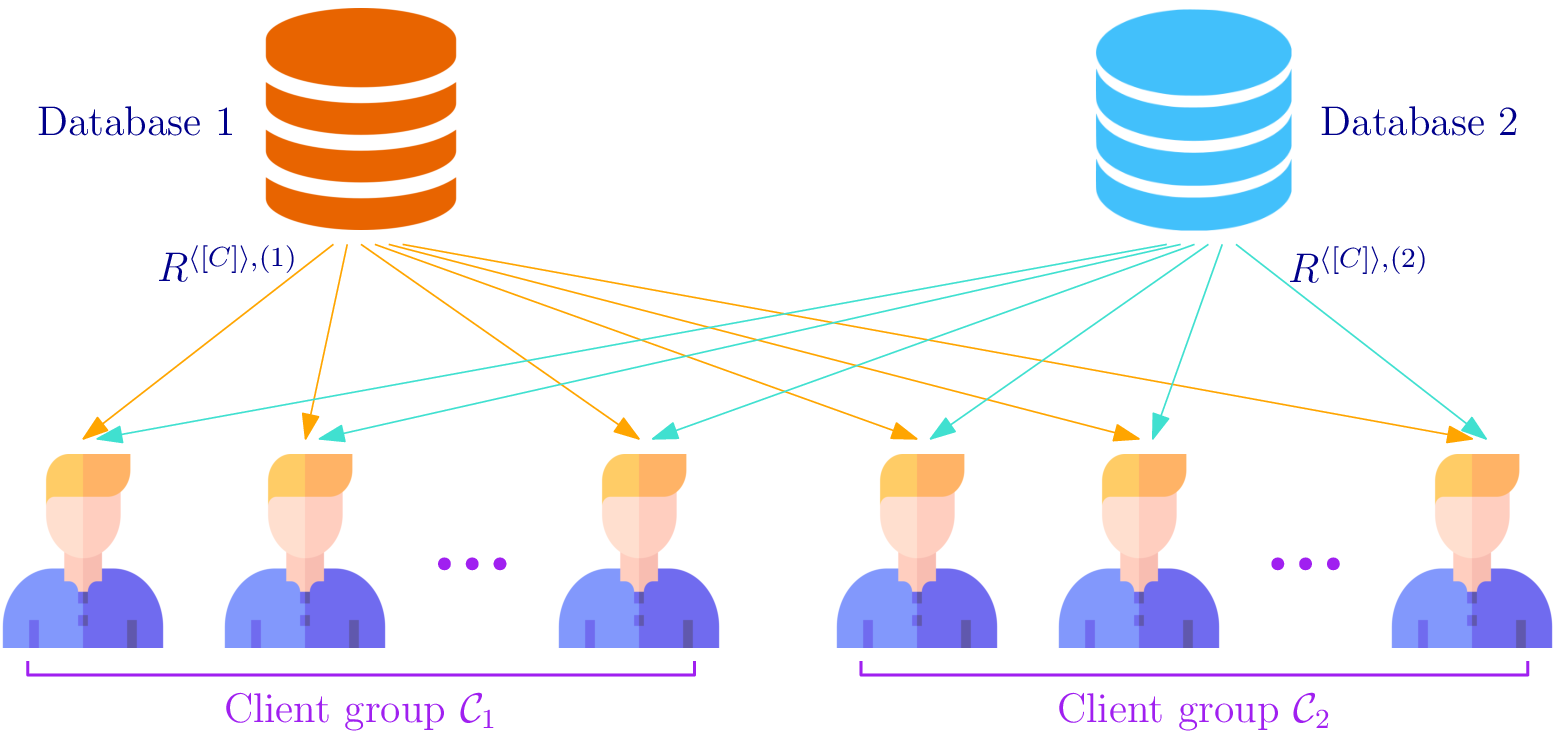}
\caption{Data flow in the common randomness allocation phase of our FSL system model.}
\label{FSL_Randomness}
\end{figure}

We start with an achievable scheme for the second type of common randomness allocation. Database $1$ in the server randomly selects a set of symbols $\{R^{(1)}_0,R^{(1)}_1,\dots,R^{(1)}_{C-1}\}$ from $\mathbb{F}_q$ under a uniform distribution and then broadcasts this set to each selected client. Likewise, database $2$ performs the same selection procedure as database $1$ and then broadcasts a set of symbols $\{R^{(2)}_0,R^{(2)}_1,\dots,R^{(2)}_{C-1}\}$ to each selected client as well. Thus, each client obtains a new set of symbols with a set length $C$ through element-wise summation within the field, 
\begin{align}
    R_i = R^{(1)}_i + R^{(2)}_i, \quad i = 0,1,\dots,C-1
\end{align}
Subsequently, each client appends one more symbol $R_C$ to the existing set such that the sum equals zero, i.e., $\sum_{i \in [C]} R_i = 0$. Since the two random sets that are transmitted from the server to each client are the same, these selected clients ultimately share a common randomness set $\{R_0,R_1,\dots,R_C\}$. Further, the individual databases in the server have no knowledge about this set because of the one-time pad encryption.

We next consider the first type of common randomness allocation, i.e., the allocation of $c$. Note that $c$ needs to be uniform in $\mathbb{F}_q \backslash \{0\}$. We could use a modified version of the above method as follows: Each database individually selects a random symbol from $\mathbb{F}_q$ under a uniform distribution and then broadcasts its selected symbol to each client. The global common randomness symbol $c$ is calculated as the sum of the two random values that are transmitted from the two databases. However, $c$ now can take the value $0$ with probability $\frac{1}{|q|}$, thus, the constraint that $c$ is uniformly distributed over $\mathbb{F}_q \backslash \{0\}$ is not immediately satisfied. The two databases in the server can repeat this procedure until $c$ falls into the allowed region, which would require feedback from the clients as explained above.  

In order to overcome this shortcoming, we propose a novel common randomness allocation method via a broadcast variation of the RSPIR scheme introduced in \cite{RSPIR}. We consider a RSPIR problem with $N = 2, K = |q| - 1, L = 1$ and make use of the potentially suboptimal\footnote{This scheme was proved to be optimal for $K=2, 3$, but is a valid scheme for any $K$.} RSPIR achievable scheme provided in \cite[Section~V]{RSPIR}. The corresponding message set stored in the two databases is set as $W_1 = 1, W_2 = 2, \dots, W_{|q|-1} = |q|-1$. We note that there is no need to protect the privacy of the remaining messages at this point since all these messages can be globally known to the clients. As a consequence, the required server-side common randomness in the original RSPIR problem \cite{RSPIR} can be discarded.\footnote{The new idea proposed here seems to be closer to the definition of random private information retrieval (RPIR) as opposed to random \emph{symmetric} private information retrieval (RSPIR) studied in \cite{RSPIR}. RPIR and RSPIR can be further studied to design more efficient and powerful common randomness construction among the clients. We leave this as an interesting future research direction.} Database $1$ has the following set of messages and broadcasts one of them randomly to active clients, 
\begin{align}
    c^{(1)}_1 &= \emptyset \\
    c^{(1)}_2 &= (W_1 + W_2, W_2 + W_3, \dots, W_{|q|-2} + W_{|q|-1}) \\
    c^{(1)}_3 &= (W_1 + W_3, W_2 + W_4, \dots, W_{|q|-2} + W_1) \\
    \vspace{-0.25em}
    &\;\; \vdots \notag \\
    \vspace{-0.25em}
    c^{(1)}_{|q|-1} &= (W_1 + W_{|q|-1}, W_2 + W_1, \dots, W_{|q|-2} + W_{|q|-3}) 
\end{align}
Similarly, database $2$ has the following set of messages and broadcasts one of them randomly to active clients,
\begin{align}
    c^{(2)}_1 &= W_1 \\
    c^{(2)}_2 &= W_2 \\
    \vspace{-0.25em}
    &\;\; \vdots \notag \\
    \vspace{-0.25em}
    c^{(2)}_{|q|-2} &= W_{|q|-2}  \\
    c^{(2)}_{|q|-1} &= W_{|q|-1} 
\end{align}
By applying the decoding procedure in the RSPIR approach in \cite{RSPIR}, all clients will be able to obtain the same global common randomness symbol $c$ that is uniformly distributed over the set $\mathbb{F}_q \backslash \{0\}$. Moreover, the obtained random symbol $c$ will be unknown to each individual database in the server due to the user privacy constraint in RSPIR \cite{RSPIR}.

\subsection{Private Set Union (FSL-PSU) Phase} \label{FSL-PSU Phase}
After the FSL-CRG phase is completed, each selected client will obtain a global common randomness symbol $c$ that is uniformly distributed over $\mathbb{F}_q \backslash \{0\}$, and a set of common randomness symbols $\{u^{\langle [C] \rangle}_k\!\!: k \in [K]\}$ in which the identity $\sum_{i \in [C]} u^{\langle i \rangle}_k = 0$ is true for all $k \in [K]$, and another set of common randomness symbols $\{u_k\!\!: k \in [K]\}$ that are all uniformly distributed over $\mathbb{F}_q$. Following our distributed FSL model in Section~\ref{FSL Formulation}, $C$ selected clients in this round of FSL process are divided into two groups $\mathcal{C}_1$ and $\mathcal{C}_2$. The clients in $\mathcal{C}_1$ send their answers to database $1$ in the server, while the clients in $\mathcal{C}_2$ send their answers to database $2$ in the server. Then, the $i$th client in $\mathcal{C}_1$, after mapping its desired submodel index set $\Gamma^{\langle i \rangle}$ into a corresponding incidence vector $Y^{\langle i \rangle}$, constructs its answer as,
\begin{align}
    A^{\langle i \rangle,(1)}_{U,1} = \{c(Y^{\langle i \rangle}_1+u^{\langle i \rangle}_1),c(Y^{\langle i \rangle}_2+u^{\langle i \rangle}_2),\dots,c(Y^{\langle i \rangle}_K+u^{\langle i \rangle}_K)\}
\end{align}
Similarly, the $i$th client in $\mathcal{C}_2$ constructs its answer as,
\begin{align}
    A^{\langle i \rangle,(2)}_{U,1} = \{c(Y^{\langle i \rangle}_1+u^{\langle i \rangle}_1),c(Y^{\langle i \rangle}_2+u^{\langle i \rangle}_2),\dots,c(Y^{\langle i \rangle}_K+u^{\langle i \rangle}_K)\}
\end{align}
Once the first database in the server receives all the answers from its associated clients in $\mathcal{C}_1$, it produces a corresponding response to be downloaded as, 
\begin{align}
    D^{\langle \theta_1 \rangle,(1)}_{U,2} = \biggl\{c\sum_{i \in \mathcal{C}_1} (Y^{\langle i \rangle}_k + u^{\langle i \rangle}_k)+S_k\!\!: k \in [K] \biggr\} 
\end{align}
where $\{S_k\!\!: k \in [K]\}$ are shared server-side common randomness symbols that are uniformly selected from $\mathbb{F}_q$ as well. This produced response $D^{\langle \theta_1 \rangle,(1)}_{U,2}$ will then be downloaded by a random client whose index $\theta_1$ belongs to $\mathcal{C}_1$. Afterwards, the $\theta_1$th client processes the received response by adding extra common randomness to it, and then, forwards the following answer to both databases in the server,
\begin{align}
    A^{\langle \theta_1 \rangle,([2])}_{U,2} = \biggl\{c\sum_{i \in \mathcal{C}_1} (Y^{\langle i \rangle}_k + u^{\langle i \rangle}_k)+u_k+S_k\!\!: k \in [K] \biggr\} 
\end{align}
Likewise, the second database produces a response to be downloaded as follows, after receiving all the answers in the first step from the client group $\mathcal{C}_2$,
\begin{align}
    D^{\langle \theta_2 \rangle,(2)}_{U,2} = \biggl\{c\sum_{i \in \mathcal{C}_2} (Y^{\langle i \rangle}_k + u^{\langle i \rangle}_k)-S_k\!\!: k \in [K] \biggr\}
\end{align}
This produced response will then be downloaded by a random client in $\mathcal{C}_2$ whose index is denoted by $\theta_2$. Afterwards, like the $\theta_1$th client, this client also forwards the following further processed answer to both databases in the server,
\begin{align}
    A^{\langle \theta_2 \rangle,([2])}_{U,2} = \biggl\{c\sum_{i \in \mathcal{C}_2} (Y^{\langle i \rangle}_k + u^{\langle i \rangle}_k)-u_k-S_k\!\!: k \in [K] \biggr\}
\end{align}
After collecting these two answer sets in the second communication round, each individual database $j$ in the server is ready to derive the desired submodel union $\Gamma$ by performing the following element-wise summation,
\begin{align}
    A^{\langle \theta_1 \rangle,(j)}_{U,2} +A^{\langle \theta_2 \rangle,(j)}_{U,2} = \biggl\{&c\sum_{i \in \mathcal{C}_1} (Y^{\langle i \rangle}_1 + u^{\langle i \rangle}_1)+u_1+S_1 + c\sum_{i \in \mathcal{C}_2} (Y^{\langle i \rangle}_1 + u^{\langle i \rangle}_1)-u_1-S_1, \notag \\
    &c\sum_{i \in \mathcal{C}_1} (Y^{\langle i \rangle}_2 + u^{\langle i \rangle}_2)+u_2+S_2+c\sum_{i \in \mathcal{C}_2} (Y^{\langle i \rangle}_2 + u^{\langle i \rangle}_2)-u_2-S_2, \notag \\
    \vspace{-0.25em}
    &\qquad \vdots \notag \\
    \vspace{-0.25em}
    &c\sum_{i \in \mathcal{C}_1} (Y^{\langle i \rangle}_K + u^{\langle i \rangle}_K)+u_K+S_K + c\sum_{i \in \mathcal{C}_2} (Y^{\langle i \rangle}_K + u^{\langle i \rangle}_K)-u_K-S_K \biggr\} \\
    = \biggl\{&c\!\sum_{i \in [C]} Y^{\langle i \rangle}_1,c\!\sum_{i \in [C]} Y^{\langle i \rangle}_2,\dots,c\!\sum_{i \in [C]} Y^{\langle i \rangle}_K\biggr\}
\end{align}

\paragraph{FSL-PSU reliability:} Each individual database $j$ in the server makes use of the $K$ elements in $A^{\langle \theta_1 \rangle,(j)}_{U,2} +A^{\langle \theta_2 \rangle,(j)}_{U,2}$ to decide whether an arbitrary element in the set $[K]$ is in the ultimate submodel index union, and thereby, to determine $\Gamma$. Let us use an arbitrary index $k$ as an example to analyze the statement above. On the one hand, if any client’s desired submodel index set includes $k$, the sum $\sum_{i \in [C]}Y^{\langle i \rangle}_k$ must be a value that is not zero and the expression $c\sum_{i \in [C]}Y^{\langle i \rangle}_k$ must be in $\mathbb{F}_q \backslash \{0\}$. On the other hand, if none of these clients' desired submodel index set includes $k$, the sum $\sum_{i \in [C]}Y^{\langle i \rangle}_k$ and its associated expression $c \sum_{i \in [C]}Y^{\langle i \rangle}_k$ are both equal to zero. Therefore, each database utilizes the value of its calculated expression $c \sum_{i \in [C]}Y^{\langle i \rangle}_k$ (whether it is zero or not) to judge whether the index $k$ is in the union $\Gamma$ or not. Following the same analysis for each $k \in [K]$, both databases can ultimately obtain the correct submodel union $\Gamma$. Thus, the FSL-PSU reliability is satisfied.

\paragraph{FSL-PSU privacy:} We analyze the FSL-PSU privacy based on the availability of the answer sets $\{A^{\langle \mathcal{C}_1 \rangle,(1)}_{U,1},A^{\langle \mathcal{C}_2 \rangle,(2)}_{U,1}\}$. For all $i \in [C]$ and all $k \in \Gamma$, the commom randomness $u^{\langle i \rangle}_k$ is used to protect the privacy of $Y^{\langle i \rangle}_k$ such that each database knows nothing about the value of $Y^{\langle i \rangle}_k$ because of the one-time pad encryption. Further, for all $k \in \Gamma$, the common randomness $c$ is used to protect the privacy of $\sum_{i \in [C]}Y^{\langle i \rangle}_k$ such that each database knows nothing about the value of $\sum_{i \in [C]}Y^{\langle i \rangle}_k$ beyond that this sum is zero or not because of the finite cyclic group under multiplication in $\mathbb{F}_q \backslash \{0\}$. Hence, the FSL-PSU privacy is preserved when the answer sets $\{A^{\langle \mathcal{C}_1 \rangle,(1)}_{U,1},A^{\langle \mathcal{C}_2 \rangle,(2)}_{U,1}\}$ are received by each database. The concrete proof follows from the proof of client's privacy in \cite[Subsection~V.B]{MP-PSI_journal}. In reality, the received answer in database $1$ is $\{A^{\langle \mathcal{C}_1 \rangle,(1)}_{U,1},A^{\langle \theta_1 \rangle,(1)}_{U,2},A^{\langle \theta_2 \rangle,(1)}_{U,2}\}$, which is equivalent to $A^{\langle \mathcal{C}_1 \rangle,(1)}_{U,1}$ and $\{c\sum_{i \in [C]}Y^{\langle i \rangle}_k\!\!:k \in [K]\}$ because of the unknown extra common randomness $\{u_k\!\!:k \in [K]\}$. Meanwhile, the received answer in database $2$ is $\{A^{\langle \mathcal{C}_2 \rangle,(2)}_{U,1},A^{\langle \theta_1 \rangle,(2)}_{U,2},A^{\langle \theta_2 \rangle,(2)}_{U,2}\}$, which is equivalent to $A^{\langle \mathcal{C}_2 \rangle,(2)}_{U,1}$ and $\{c\sum_{i \in [C]}Y^{\langle i \rangle}_k\!\!:k \in [K]\}$ for the same reason. That means that each database receives less information with respect to the incidence vectors $Y^{\langle [C] \rangle}$ than the answer set $\{A^{\langle \mathcal{C}_1 \rangle,(1)}_{U,1},A^{\langle \mathcal{C}_2 \rangle,(2)}_{U,1}\}$. Thus, the FSL-PSU privacy constraint is satisfied.

\paragraph{FSL-PSU inter-client privacy:} Only the clients $\theta_1$ and $\theta_2$ receive information from outside. Due to the existence of the unknown server-side common randomness in the downloads $D^{\langle \theta_1 \rangle,(1)}_{U,2}$ and $D^{\langle \theta_2 \rangle,(2)}_{U,2}$, neither the $\theta_1$th client nor the $\theta_2$th client can learn any knowledge about the incidence vector within the other clients. Therefore, the FSL-PSU inter-client privacy constraint is satisfied as well.

\paragraph{FSL-PSU communication cost:} Without considering the communication cost generated in the common randomness generation phase, the communication cost in this phase is $(C+6)K$ in $q$-ary bits. Moreover, following the common randomness generation approach provided in Section~\ref{Common Randomness Allocation Phase}, the additional communication cost is $2CK$ for the common randomness sets $\{u^{\langle [C] \rangle}_k\!\!: k \in [K]\}$ and $\{u_k\!\!: k \in [K]\}$. Further, for the global common randomness symbol $c$, the required communication cost is approximately $2(|q|-1)C$ in $q$-ary bits, which is negligible since the value of $K$ is generally very large. Therefore, the total communication cost in this phase is $(3C+6)K$ in $q$-ary bits.

\paragraph{FSL-PSU client drop-out robustness:} In the first step of FSL-PSU phase, for all $i \in [C]$, client $i$ sends its generated answer to its associated database in the server. Without loss of generality, we assume that a set of clients $\mathcal{C}_{1,\mathcal{D}}$ belonging to the first client group $\mathcal{C}_1$ and another set of clients $\mathcal{C}_{2,\mathcal{D}}$ belonging to the second client group $\mathcal{C}_2$ drop out in this step. Hence, the response to be downloaded produced by database $1$ is as follows, 
\begin{align}
    D^{\prime\langle \theta_1 \rangle,(1)}_{U,2} = \biggl\{c\!\!\!\!\sum_{i \in \mathcal{C}_1 \backslash \mathcal{C}_{1,\mathcal{D}}} \!\!\! (Y^{\langle i \rangle}_k + u^{\langle i \rangle}_k)+S_k\!\!: k \in [K] \biggr\} 
\end{align}
After receiving the response $D^{\prime\langle \theta_1 \rangle,(1)}_{U,2}$ as well as the index set of out-of-operation clients $\mathcal{C}_{1,\mathcal{D}}$, client $\theta_1$ can adjust the answer by additionally appending the sum of missing common randomness symbols $\sum_{i \in \mathcal{C}_{1,\mathcal{D}}} u^{\langle i \rangle}_k$ for all $k \in [K]$. Hence, the answer generated by client $\theta_1$ in the second step is as follows,
\begin{align}
    A^{\prime\langle \theta_1 \rangle,([2])}_{U,2} = \biggl\{c\!\!\!\! \sum_{i \in \mathcal{C}_1 \backslash \mathcal{C}_{1,\mathcal{D}}} \!\!\!\!\! Y^{\langle i \rangle}_k + c \sum_{i \in \mathcal{C}_1} u^{\langle i \rangle}_k+u_k+S_k\!\!: k \in [K] \biggr\} 
\end{align}
Likewise, the response to be downloaded produced by database $2$ is as follows, 
\begin{align}
    D^{\prime\langle \theta_2 \rangle,(2)}_{U,2} = \biggl\{c\!\!\!\!\sum_{i \in \mathcal{C}_2 \backslash \mathcal{C}_{2,\mathcal{D}}} \!\!\! (Y^{\langle i \rangle}_k + u^{\langle i \rangle}_k)+S_k\!\!: k \in [K] \biggr\} 
\end{align}
The answer generated by client $\theta_2$ in the second step is as follows,
\begin{align}
    A^{\prime\langle \theta_2 \rangle,([2])}_{U,2} = \biggl\{c\!\!\!\! \sum_{i \in \mathcal{C}_2 \backslash \mathcal{C}_{2,\mathcal{D}}} \!\!\!\!\! Y^{\langle i \rangle}_k + c \sum_{i \in \mathcal{C}_2} u^{\langle i \rangle}_k-u_k-S_k\!\!: k \in [K] \biggr\} 
\end{align}
After collecting the answers $A^{\prime\langle \theta_1 \rangle,(j)}_{U,2}$ and $A^{\prime\langle \theta_2 \rangle,(j)}_{U,2}$, by adding them up element-wisely, each individual database $j$ in the server will obtain the union result as $\Gamma_{\mathcal{C}_{1,\mathcal{D}}} \cup \Gamma_{\mathcal{C}_{2,\mathcal{D}}}$ containing all the active selected clients following the steps in the FSL-SPU reliability constraints. Another non-trivial point is that the randomly selected clients $\theta_1$ and $\theta_2$ may also drop-out during the implementation of FSL-PSU phase step 2. In a practical application, a potential solution is that each database individually randomly selects a small set of clients to route the information in parallel like the clients $\theta_1$ and $\theta_2$. Further, we may use the observed client drop-out rate to determine the cardinality of this small relaying set. 

\paragraph{FSL-PSU client late-arrival robustness:} Without loss of generality, we assume that an answer generated by an arbitrary client with index $i \in \mathcal{C}_j$ in the first step arrives at database $j$ late. Even though database $j$ receives the information $A^{\langle i \rangle,(j)}_{U,1}$ separately, it is still not able to extract any information about the incidence vector $Y^{\langle i \rangle}$ from the received answers in the two steps of FSL-PSU phase because of the unknown extra common randomness $\{u_k\!\!: k \in [K]\}$. This conclusion can be extended to a set of arbitrary clients who arrive at the same database late. Moreover, it is easy to guarantee that this late answer will never be transmitted to any other client in order to avoid information leakage.

\paragraph{FSL-PSU database drop-out robustness:} If database $1$ drops-out and cannot function normally, database $2$ can still receive the answers $A^{\langle \mathcal{C}_2 \rangle,(2)}_{U,1}$ and $A^{\langle \theta_2 \rangle,(2)}_{U,2}$ as normal but cannot receive any answer from the relaying client in $\mathcal{C}_1$. In order to derive the union $\Gamma_{\mathcal{C}_2}$ through decoding the set $\{c\sum_{i \in \mathcal{C}_2} Y^{\langle i \rangle}_k\!\!: k \in [K]\}$ from its existing information, database $2$ needs to communicate with client $\theta_2$ one more time for the sake of the values of $\{c\sum_{i \in \mathcal{C}_1} u^{\langle i \rangle}_k \!\!: k \in [K]\}$. Likewise, if database $2$ cannot function normally, this time, database $1$ can still derive the union $\Gamma_{\mathcal{C}_1}$ following the same way. Further, if we encounter client drop-out or client late-arrival in addition to the occurrence of database drop-out, the last two robustness analyses can be utilized accordingly to make this scheme function well.

\subsection{Private Write (FSL-write) Phase} \label{FSL-write Phase}
When the FSL-PSU phase is complete, the server learns the desired submodel union $\Gamma$ from all the selected clients in this FSL round. Then, each database in the server individually sends the set of submodels $M_\Gamma$ to its associated clients. From the FSL-CRG phase preceding the FSL-PSU phase, each selected client has also obtained two sets of common randomness symbols $\{w^{\langle [C] \rangle}_{k,l}\!\!: k \in \Gamma, l \in [L]\}$ and $\{w_{k,l}\!\!: k \in \Gamma, l \in [L]\}$ from $\mathbb{F}_q$. Likewise, we always have $\sum_{i \in [C]} w^{\langle i \rangle}_{k,l} = 0$ for all $k \in \Gamma$ and $l \in [L]$. Therefore, the $i$th client in $\mathcal{C}_1$ will generate the increments for each desired submodel whose index belongs to $\Gamma^{\langle i \rangle}$ according to its local training data and then construct a well-processed answer accordingly. Specifically, for all $k \in \Gamma^{\langle i \rangle}$, the answer symbols are generated in the following form, 
\begin{align}
    A^{\langle i \rangle,(1)}_{W,1}(k) = \{\Delta^{\langle i \rangle}_{k,1} + w^{\langle i \rangle}_{k,1}, \Delta^{\langle i \rangle}_{k,2} + w^{\langle i \rangle}_{k,2}, \dots, \Delta^{\langle i \rangle}_{k,L} + w^{\langle i \rangle}_{k,L}\}
\end{align}
In addition, for all $k \in \Gamma \backslash \Gamma^{\langle i \rangle}$, the answer symbols are generated as follows without any updates concerning the current submodel,
\begin{align}
    A^{\langle i \rangle,(1)}_{W,1}(k) = \{w^{\langle i \rangle}_{k,1}, w^{\langle i \rangle}_{k,2}, \dots, w^{\langle i \rangle}_{k,L}\}
\end{align}
Thus, the ultimate answer generated by this client in the first step is $A^{\langle i \rangle,(1)}_{W,1} = \{A^{\langle i \rangle,(1)}_{W,1}(k)\!\!: k \in \Gamma\}$. The $i$th client in $\mathcal{C}_2$ will generate an ultimate answer $A^{\langle i \rangle,(2)}_{W,1} = \{A^{\langle i \rangle,(2)}_{W,1}(k)\!\!: k \in \Gamma\}$ in the same way, where
\begin{align}
    A^{\langle i \rangle,(2)}_{W,1}(k) = \{&\Delta^{\langle i \rangle}_{k,1} + w^{\langle i \rangle}_{k,1}, \Delta^{\langle i \rangle}_{k,2} + w^{\langle i \rangle}_{k,2}, \dots, \Delta^{\langle i \rangle}_{k,L} + w^{\langle i \rangle}_{k,L}\}, \quad k \in \Gamma^{\langle i \rangle} \\
    A^{\langle i \rangle,(2)}_{W,1}(k) = \{&w^{\langle i \rangle}_{k,1}, w^{\langle i \rangle}_{k,2}, \dots, w^{\langle i \rangle}_{k,L}\}, \quad k \in \Gamma \backslash \Gamma^{\langle i \rangle}
\end{align}
Subsequently, each client sends its answer to its associated database in the server. These two databases also share another set of server-side common randomness symbols $\{S_{k,l}\!\!: k \in [K], l \in [L]\}$ from $\mathbb{F}_q$. Let $C^{(1)}_k$ be the index set of clients in the first client group $\mathcal{C}_1$ whose desired submodel index set includes the index $k$, i.e., $C^{(1)}_k = \{i \in \mathcal{C}_1| k \in \Gamma^{\langle i \rangle}\}$. Similarly, $C^{(2)}_k$ and $C_k$ are defined as $\{i \in \mathcal{C}_2| k \in \Gamma^{\langle i \rangle}\}$ and $\{i \in [C]|k \in \Gamma^{\langle i \rangle}\}$, respectively. After collecting all the answers $A^{\langle \mathcal{C}_1 \rangle,(1)}_{W,1}$ from $\mathcal{C}_1$, database $1$ calculates the following aggregation increment for the $l$th symbol of the $k$th submodel where $k$ belongs to the union set $\Gamma$, 
\begin{align}
     \sum_{i \in C^{(1)}_k} \Bigl(\Delta^{\langle i \rangle}_{k,l} + w^{\langle i \rangle}_{k,l}\Bigl) + \sum_{i \in \mathcal{C}_1 \backslash C^{(1)}_k} \!\! w^{\langle i \rangle}_{k,l} = \sum_{i \in C^{(1)}_k}\!\! \Delta^{\langle i \rangle}_{k,l} + \sum_{i \in \mathcal{C}_1}  w^{\langle i \rangle}_{k,l}
\end{align}
As in the last FSL-PSU phase, after adding server-side common randomness, the corresponding response is produced as follows and will be downloaded by the client $\theta_1$, 
\begin{align}
    D^{\langle \theta_1 \rangle,(1)}_{W,2} = \biggl\{\sum_{i \in C^{(1)}_k}\!\! \Delta^{\langle i \rangle}_{k,l} + \sum_{i \in \mathcal{C}_1}  w^{\langle i \rangle}_{k,l} + S_{k,l} \!\!: k \in \Gamma, l \in [L]\biggr\}
\end{align}
Once this response is received by client $\theta_1$, this client only adds extra common randomness and then forwards the following answer to both databases,
\begin{align}
    A^{\langle \theta_1 \rangle,([2])}_{W,2} = \biggl\{\sum_{i \in C^{(1)}_k}\!\! \Delta^{\langle i \rangle}_{k,l} + \sum_{i \in \mathcal{C}_1}  w^{\langle i \rangle}_{k,l} + w_{k,l} + S_{k,l} \!\!: k \in \Gamma, l \in [L]\biggr\}
\end{align}
Meanwhile, after collecting all the answers $A^{\langle \mathcal{C}_2 \rangle,(2)}_{W,1}$ from $\mathcal{C}_2$, database $2$ produces the following response and this response will be downloaded by the client $\theta_2$,
\begin{align}
    D^{\langle \theta_2 \rangle,(2)}_{W,2} = \biggl\{\sum_{i \in C^{(2)}_k}\!\! \Delta^{\langle i \rangle}_{k,l} + \sum_{i \in \mathcal{C}_2}  w^{\langle i \rangle}_{k,l} - S_{k,l} \!\!: k \in \Gamma, l \in [L]\biggr\}
\end{align}
The answer that is forwarded by client $\theta_2$ to both databases is as follows,
\begin{align}
    A^{\langle \theta_2 \rangle,([2])}_{W,2} = \biggl\{\sum_{i \in C^{(2)}_k}\!\! \Delta^{\langle i \rangle}_{k,l} + \sum_{i \in \mathcal{C}_2}  w^{\langle i \rangle}_{k,l} - w_{k,l} - S_{k,l} \!\!: k \in \Gamma, l \in [L]\biggr\}
\end{align}

At this point, each individual database in the server is ready to aggregate the updates as desired from all the selected clients in this round of FSL. For the $l$th symbol of the $k$th submodel in $M_{\Gamma}$, the ultimate aggregation increment is calculated as follows,
\begin{align}
    \sum_{i \in C^{(1)}_k}\!\! &\Delta^{\langle i \rangle}_{k,l} + \sum_{i \in \mathcal{C}_1}  w^{\langle i \rangle}_{k,l} + w_{k,l} + S_{k,l} + \sum_{i \in C^{(2)}_k}\!\! \Delta^{\langle i \rangle}_{k,l} + \sum_{i \in \mathcal{C}_2}  w^{\langle i \rangle}_{k,l} - w_{k,l} - S_{k,l} \nonumber\\
    &= \sum_{i \in C^{(1)}_k \cup C^{(2)}_k}\!\! \Delta^{\langle i \rangle}_{k,l} + \sum_{i \in \mathcal{C}_1 \cup \mathcal{C}_2}  w^{\langle i \rangle}_{k,l} \\ 
    &= \sum_{i \in C_{k}} \Delta^{\langle i \rangle}_{k,l}
\end{align}
The updated $l$th symbol of the $k$th submodel $M^{\prime}_{k,l}$ stored in the server after performing this round of FSL-write should finally be 
\begin{align}
    M^{\prime}_{k,l} = M_{k,l} + \sum_{i \in C_k} \Delta^{\langle i \rangle}_{k,l}
\end{align}

It is clear that this scheme satisfies the FSL-write reliability constraint. It is important to note that the scheme in FSL-write phase is essentially a repetitive application of a simplified version of the scheme in FSL-PSU phase without involving the global common randomness symbol $c$. Thus, the FSL-write scheme satisfies the FSL-write privacy constraint as well as FSL-write inter-client privacy constraints, and also is robust against client drop-out, client late-arrival and database drop-out events.

\paragraph{FSL-write communication cost:} If we also do not consider the communication cost generated in the accompanying common randomness generation, the communication cost is $(2C+6) |\Gamma|L$ in $q$-ary bits in which $C|\Gamma|L$ is for the clients to download the submodels from the server. The communication cost of obtaining the common randomness sets is $\{w^{\langle [C] \rangle}_{k,l}\!\!: k \in \Gamma, l \in [L]\}$ and $\{w_{k,l}\!\!: k \in \Gamma, l \in [L]\}$ for clients is $2C|\Gamma|L$ in $q$-ary bits. Therefore, the total communication cost in this phase is $(4C+6)|\Gamma|L$ in $q$-ary bits.

\vspace{2em}
The complete procedure involving FSL-PSU phase and FSL-write phase in this round of FSL process can be executed repeatedly to update the full learning model iteratively until a pre-specified termination criterion is met. All the characteristics introduced above are preserved in all FSL rounds.

\section{Conclusion and Discussion}

We proposed a new private distributed FSL achievable scheme with a communication cost that is order-wise similar to the communication cost of existing schemes which provide much weaker privacy guarantees. Compared to the existing schemes with similar privacy guarantees, our proposed scheme does not require noisy storage of the submodels in the databases. Our proposed scheme is resilient against client drop-outs, client late-arrivals, and database drop-outs. The main ideas of this scheme are based on private set union (PSU) and its variation for private-write, together with random private information retrieval and one-time pads for needed common randomness generation at the client side. 

Our scheme starts with replicated storage of the submodels in two non-colluding databases at the server together with some amount of server-side common randomness. Our scheme privately generates needed common randomness at the client side, privately determines the union of the indices of the submodels to be updated, and privately writes the updated submodels back to the databases. Neither the indices of the submodels updated within the union, nor their updated values are leaked to the databases.  

In this work, we considered the simplest version of this new formulation. The issues that need to be studied further include: 1) The case when the server has more than two databases. 2) Privacy of stored data against databases. 3) Colluding databases. 4) Byzantine databases that send erroneous information. 5) Collusion among clients. 6) Byzantine clients that send erroneous updates to poison the learning process. 7) Schemes to reduce the communication and storage cost, and potential communication-storage trade-off. 8) MDS coded storage and/or MDS coded user-side common randomness. 9) Optimum partitioning of the clients among databases, especially, with colluding databases under a known colluding structure.   

\bibliographystyle{unsrt}
\bibliography{Journal2022}

\begin{thebibliography}{10}

\bibitem{FL}
B.~McMahan and D.~Ramage.
\newblock Federated learning: Collaborative machine learning without
  centralized training data.
\newblock Available at
  https://ai.googleblog.com/2017/04/federated-learning-collaborative.html.

\bibitem{FL_Yangconcept}
Q.~Yang, Y.~Liu, T.~Chen, and Y.~Tong.
\newblock Federated machine learning: Concept and applications.
\newblock {\em ACM Transactions on Intelligent Systems and Technology},
  10(2):1--19, March 2019.

\bibitem{SecAgg}
K.~Bonawitz, V.~Ivanov, et~al.
\newblock Practical secure aggregation for privacy preserving machine learning.
\newblock In {\em Proceedings of the 2017 ACM SIGSAC Conference on Computer and
  Communications Security}, page 1175–1191, 2017.

\bibitem{SecAgg+}
J.~Bell, K.~A. Bonawitz, A.~Gascón, T.~Lepoint, and M.~Raykova.
\newblock Secure single-server aggregation with (poly)logarithmic overhead.
\newblock In {\em Cryptology ePrint Archive}, 2020.

\bibitem{TurboAgg}
J.~So, B.~Guler, and A.~S. Avestimehr.
\newblock Turbo-aggregate: Breaking the quadratic aggregation barrier in secure
  federated learning.
\newblock In {\em Cryptology ePrint Archive}, 2020.

\bibitem{FastSecAgg}
S.~Kadhe and K.~Ramchandran N.~Rajaraman~abd O.~O.~Koyluoglu.
\newblock Fastsecagg: Scalable secure aggregation for privacy-preserving
  federated learning.
\newblock Available at arXiv:2009.11248.

\bibitem{IT_SecAgg}
Y.~Zhao and H.~Sun.
\newblock Information theoretic secure aggregation with user dropouts.
\newblock {\em IEEE Trans. on Info. Theory}, 68(11):7471--7484, November 2022.

\bibitem{IT_SecAgg_UGKey}
K.~Wan, H.~Sun, M.~Ji, and G.~Caire.
\newblock Information theoretic secure aggregation with uncoded groupwise keys.
\newblock Available at arXiv:2204.11364.

\bibitem{IT_SecAgg_Region}
Y.~Zhao and H.~Sun.
\newblock Secure summation: Capacity region, groupwise key, and feasibility.
\newblock Available at arXiv:2205.08458.

\bibitem{LightSecAgg}
J.~So, C.~He, et~al.
\newblock Lightsecagg: a lightweight and versatile design for secure
  aggregation in federated learning.
\newblock In {\em Proceedings of Machine Learning and Systems}, pages 694--720,
  2022.

\bibitem{FSL}
C.~Niu, F.~Wu, S.~Tang, L.~Hua, R.~Jia, C.~Lv, Z.~Wu, and G.~Chen.
\newblock Billion-scale federated learning on mobile clients: A submodel design
  with tunable privacy.
\newblock In {\em Proceedings of the 26th Annual International Conference on
  Mobile Computing and Networking}, pages 1--14, 2020.

\bibitem{KS05}
L.~Kissner and D.~Song.
\newblock Privacy-preserving set operations.
\newblock In {\em Advances in Cryptology -- CRYPTO 2005}, pages 241--257, 2005.

\bibitem{PSU}
K.~Frikken.
\newblock Privacy-preserving set union.
\newblock In {\em Applied Cryptography and Network Security}, pages 237--252,
  2007.

\bibitem{PMG_Agg}
C.~Naim, R.~G.~L. D’Oliveira, and S.~El Rouayheb.
\newblock Private multi-group aggregation.
\newblock {\em IEEE Jour. on Selected Areas in Commun.}, 40(3):800--814, March
  2022.

\bibitem{XSTFSL}
Z.~Jia and S.~A. Jafar.
\newblock {$X$}-secure {$T$}-private federated submodel learning with elastic
  dropout resilience.
\newblock {\em IEEE Trans. on Info. Theory}, 68(8):5418--5439, August 2022.

\bibitem{XSTPIR}
Z.~Jia, H.~Sun, and S.~A. Jafar.
\newblock Cross subspace alignment and the asymptotic capacity of ${X}$-secure
  ${T}$-private information retrieval.
\newblock {\em IEEE Trans. on Info. Theory}, 65(9):5783--5798, September 2019.

\bibitem{Sajani_FSL1}
S.~Vithana and S.~Ulukus.
\newblock Efficient private federated submodel learning.
\newblock In {\em IEEE ICC}, pages 3394--3399, May 2022.

\bibitem{Sajani_FSL_Trans}
S.~Vithana and S.~Ulukus.
\newblock Private read update write ({PRUW}) in federated submodel learning
  ({FSL}): Communication efficient schemes with and without sparsification.
\newblock Available at arXiv:2209.04421.

\bibitem{PIR_ORI}
B.~Chor, E.~Kushilevitz, O.~Goldreich, and M.~Sudan.
\newblock Private information retrieval.
\newblock {\em Journal of the ACM}, 45(6):965--981, November 1998.

\bibitem{SPIR_ORI}
Y.~Gertner, Y.~Ishai, E.~Kushilevitz, and T.~Malkin.
\newblock Protecting data privacy in private information retrieval schemes.
\newblock {\em Journal of Computer and System Sciences}, 60(3):592--629, June
  2000.

\bibitem{PIR}
H.~Sun and S.~A. Jafar.
\newblock The capacity of private information retrieval.
\newblock {\em IEEE Trans. on Info. Theory}, 63(7):4075--4088, July 2017.

\bibitem{PIR_coded}
K.~Banawan and S.~Ulukus.
\newblock The capacity of private information retrieval from coded databases.
\newblock {\em IEEE Trans. on Info. Theory}, 64(3):1945--1956, March 2018.

\bibitem{ColludingPIR}
H.~Sun and S.~A. Jafar.
\newblock The capacity of robust private information retrieval with colluding
  databases.
\newblock {\em IEEE Trans. on Info. Theory}, 64(4):2361--2370, April 2018.

\bibitem{BPIRjournal}
K.~Banawan and S.~Ulukus.
\newblock The capacity of private information retrieval from {B}yzantine and
  colluding databases.
\newblock {\em IEEE Trans. on Info. Theory}, 65(2):1206--1219, February 2019.

\bibitem{PIR_Eavesdropper}
Q.~Wang, H.~Sun, and M.~Skoglund.
\newblock The capacity of private information retrieval with eavesdroppers.
\newblock {\em IEEE Trans. on Info. Theory}, 65(5):3198--3214, May 2019.

\bibitem{MM-PIR}
K.~Banawan and S.~Ulukus.
\newblock Multi-message private information retrieval: Capacity results and
  near-optimal schemes.
\newblock {\em IEEE Trans. on Info. Theory}, 64(10):6842--6862, October 2018.

\bibitem{tandon_cache_2017}
R.~Tandon.
\newblock The capacity of cache aided private information retrieval.
\newblock In {\em Allerton Conference}, pages 1078--1082, October 2017.

\bibitem{PrefetchingPIR}
Y.-P. Wei, K.~Banawan, and S.~Ulukus.
\newblock Fundamental limits of cache-aided private information retrieval with
  unknown and uncoded prefetching.
\newblock {\em IEEE Trans. on Info. Theory}, 65(5):3215--3232, May 2019.

\bibitem{PIR_PSI}
Z.~Chen, Z.~Wang, and S.~A. Jafar.
\newblock The capacity of ${T}$-private information retrieval with private side
  information.
\newblock {\em IEEE Trans. on Info. Theory}, 66(8):4761--4773, August 2020.

\bibitem{StorageConstrainedPIR}
M.~A. Attia, D.~Kumar, and R.~Tandon.
\newblock The capacity of private information retrieval from uncoded storage
  constrained databases.
\newblock {\em IEEE Trans. on Info. Theory}, 66(11):6617--6634, November 2020.

\bibitem{StorageCost}
C.~Tian.
\newblock On the storage cost of private information retrieval.
\newblock {\em IEEE Trans. on Info. Theory}, 66(12):7539--7549, December 2020.

\bibitem{ProbPIR}
C.~Tian, H.~Sun, and J.~Chen.
\newblock Capacity-achieving private information retrieval codes with optimal
  message size and upload cost.
\newblock {\em IEEE Trans. on Info. Theory}, 65(11):7613--7627, November 2019.

\bibitem{AleakyPIR}
I.~Samy, M.~Attia, R.~Tandon, and L.~Lazos.
\newblock Asymmetric leaky private information retrieval.
\newblock {\em IEEE Trans. on Info. Theory}, 67(8):5352--5369, August 2021.

\bibitem{SemanticPIR}
S.~Vithana, K.~Banawan, and S.~Ulukus.
\newblock Semantic private information retrieval.
\newblock {\em IEEE Trans. on Info. Theory}, 68(4):2635--2652, April 2022.

\bibitem{SPIR}
H.~Sun and S.~A. Jafar.
\newblock The capacity of symmetric private information retrieval.
\newblock {\em IEEE Trans. on Info. Theory}, 65(1):322--329, January 2019.

\bibitem{SPIR_atPIR}
Z.~Wang and S.~Ulukus.
\newblock Symmetric private information retrieval at the private information
  retrieval rate.
\newblock {\em IEEE Jour. on Selected Areas in Info. Theory}, 3(2):350--361,
  June 2022.

\bibitem{CommCost_ISIT2022}
Z.~Wang and S.~Ulukus.
\newblock Communication cost of two-database symmetric private information
  retrieval: A conditional disclosure of multiple secrets perspective.
\newblock In {\em IEEE ISIT}, pages 402--407, June 2022.

\bibitem{SPIR_Eavesdropper}
Q.~Wang and M.~Skoglund.
\newblock On {PIR} and symmetric {PIR} from colluding databases with
  adversaries and eavesdroppers.
\newblock {\em IEEE Trans. on Info. Theory}, 65(5):3183--3197, May 2019.

\bibitem{SPIR_coded}
Q.~Wang and M.~Skoglund.
\newblock Symmetric private information retrieval from {MDS} coded distributed
  storage with non-colluding and colluding servers.
\newblock {\em IEEE Trans. on Info. Theory}, 65(8):5160--5175, August 2019.

\bibitem{SPIR_Collusion}
J.~Cheng, N.~Liu, W.~Kang, and Y.~Li.
\newblock The capacity of symmetric private information retrieval under
  arbitrary collusion and eavesdropping patterns.
\newblock {\em IEEE Trans. on Info. Forensics and Security}, 17:3037--3050,
  August 2022.

\bibitem{PSI_journal}
Z.~Wang, K.~Banawan, and S.~Ulukus.
\newblock Private set intersection: A multi-message symmetric private
  information retrieval perspective.
\newblock {\em IEEE Trans. on Info. Theory}, 68(3):2001--2019, March 2022.

\bibitem{MP-PSI_journal}
Z.~Wang, K.~Banawan, and S.~Ulukus.
\newblock Multi-party private set intersection: An information-theoretic
  approach.
\newblock {\em IEEE Jour. on Selected Areas in Info. Theory}, 2(1):366--379,
  March 2021.

\bibitem{RSPIR}
Z.~Wang and S.~Ulukus.
\newblock Digital blind box: Random symmetric private information retrieval.
\newblock In {\em IEEE ITW}, pages 95--100, November 2022.

\bibitem{Prio}
H.~Corrigan-Gibbs and D.~Boneh.
\newblock Prio: Private, robust, and scalable computation of aggregate
  statistics.
\newblock In {\em Proceedings of the 14th USENIX Conference on Networked
  Systems Design and Implementation}, page 259–282, 2017.

\bibitem{DoubleBlind_PIR}
Y.~Lu, Z.~Jia, and S.~A. Jafar.
\newblock Double blind {T}-private information retrieval.
\newblock {\em IEEE Jour. on Selected Areas in Info. Theory}, 2(1):428--440,
  March 2021.

\bibitem{MultiBlind_SPIR}
J.~Zhu, Q.~Yan, and X.~Tang.
\newblock Multi-user blind symmetric private information retrieval from coded
  servers.
\newblock {\em IEEE Jour. on Selected Areas in Commun.}, 40(3):815--831, March
  2022.

\bibitem{Kim_FSL}
M.~Kim and J.~Lee.
\newblock Information-theoretic privacy in federated submodel learning.
\newblock Available at arXiv:2008.07656.

\bibitem{Sajani_FSL2}
S.~Vithana and S.~Ulukus.
\newblock Private read update write ({PRUW}) with storage constrained
  databases.
\newblock In {\em IEEE ISIT}, pages 2391--2396, June 2022.

\bibitem{Sajani_FSL3}
S.~Vithana and S.~Ulukus.
\newblock Private federated submodel learning with sparsification.
\newblock In {\em IEEE ITW}, pages 410--415, November 2022.

\bibitem{Sajani_FSL4}
S.~Vithana and S.~Ulukus.
\newblock Rate distortion tradeoff in private read update write in federated
  submodel learning.
\newblock In {\em Allerton Conference}, October 2022.

\end{thebibliography}

\end{document}